\documentclass [10pt] {article}

\usepackage{amsmath}
\usepackage{amssymb}
\usepackage{amsthm}
\usepackage{epsfig}
\usepackage{named}
\usepackage{mathrsfs}
\usepackage{fancyheadings}
\usepackage{makeidx}

\renewcommand{\baselinestretch}{1.2}

\theoremstyle{plain}
\newtheorem{proposition}{Proposition}[subsection]

\numberwithin{figure}{subsection}
\numberwithin{proposition}{subsection}

\makeindex

\begin{document}
\title{\huge{\textbf{Classical Relativity Theory}}\thanks{I am grateful to Jeremy Butterfield, Erik Curiel, and  John Earman for comments on earlier drafts.} \\  \vspace{.1in}
\normalsize{Version 2.4} 
}

\author{{David B. Malament} \\ \normalsize{Department of Logic and Philosophy of Science} \\ \normalsize{3151 Social Science Plaza} \\ \normalsize{University of
California, Irvine} \\ \normalsize{Irvine, CA\ 92697-5100} \\ \normalsize{dmalamen@uci.edu}\\
 } 
\date{}

\maketitle

\thispagestyle{fancy}
\chead{\large{To appear in: \textit{Handlbook of the Philosophy of Physics},  \\ eds. J. Butterfield  and J. Earman, Elsevier}\vspace{.1in}}

\begin{abstract}
The essay that follows is divided into two parts. In the first (section 2), I give a brief account of the structure of  classical relativity theory. In the second (section 3), I discuss three special topics:  (i) the status of the relative simultaneity relation in the context of Minkowski spacetime; (ii)  the ``geometrized" version of Newtonian gravitation theory (also known as Newton-Cartan theory); and (iii) the possibility of recovering the global geometric structure of spacetime from its ``causal structure".  

\vspace{1em}

\end{abstract}

\newpage
\tableofcontents

\newpage

\section{Introduction}

The essay that follows is divided into two parts. In the first, I give a brief account of the structure of  classical relativity theory.\footnote{I speak of  ``classical" relativity theory because considerations involving quantum mechanics will play no role. In particular, there will be no discussion of quantum field theory in curved spacetime, or of attempts to formulate a quantum theory of gravitation. (For the latter, see Rovelli (this volume,  chapter 12).)}   In the second, I discuss three special topics. 

My account in the first part (section 2) is limited in several respects. I do not discuss the historical development of classical relativity theory, nor the evidence we have for it.   I do not treat ``special relativity" as a theory in its own right that is superseded by ``general relativity".  And I do not describe known exact solutions to Einstein's equation.  (This list could be continued at great length.\footnote{Two important topics that I do not consider figure centrally in other contributions to this volume, namely the initial value formulation of relativity theory (Earman, chapter 15), and the Hamilton\-ian formulation of relativity theory (Belot, chapter 2).}) 
Instead, I  limit myself to a few fundamental ideas, and present them as clearly and precisely as I can. The account presupposes a good understanding of  basic differential geometry, and at least passing acquaintance with relativity theory itself.\footnote{A review of the needed differential geometry (and ``abstract-index notation" that I use) can be found, for example, in Wald \shortcite{Wald} and Malament \shortcite{Malament/DG}. (Some topics are also reviewed in sections 3.1 and 3.2 of Butterfield (this volume, chapter 1).)  In preparing part 1, I have drawn heavily on a number of sources. At the top of the list are  Geroch \shortcite{Geroch/GRNotes},  
Hawking and Ellis \shortcite{Hawking-Ellis},  O'Neill \shortcite{O'Neill}, Sachs and Wu  \shortcite{Sachs-Wu/article,Sachs-Wu/book}, and Wald \shortcite{Wald}.}  

 In section 3, I  first consider the status of the relative simultaneity relation in the context of Minkowski spacetime.  At issue is whether the standard relation, the one picked out by  Einstein's ``definition" of simultaneity, is conventional in character, or is rather in some significant sense forced on us.  Then I  describe the ``geometrized" version of Newtonian gravitation theory (also known as Newton-Cartan theory).  It is included here because it helps to clarify what is and is not distinctive about classical relativity theory. Finally, I consider to what extent the global geometric structure of spacetime can be recovered from its ``causal structure".\footnote{Further discussion of the foundations of classical relativity theory, from a slightly different point of view, can be found in Rovelli (this volume,  chapter 12).}

\section{The Structure of Relativity Theory}

\subsection{Relativistic  Spacetimes} \label{Relativistic  Spacetimes}

Relativity theory determines a class of geometric models for the spacetime structure of our universe (and subregions thereof such as, for example, our solar system).  Each represents a possible world (or world-region)  compatible with the constraints of the theory. 
It is convenient to describe these models in stages. We start by characterizing a broad class of ``relativistic spacetimes", and discussing their interpretation.  Later we introduce  further restrictions involving global spacetime structure and  Einstein's equation.  

We take a \emph{relativistic spacetime} to be a pair $(M, g_{ab})$, where $M$ is a smooth, connected, four-dimensional manifold, and $g_{ab}$ is a smooth, semi-Riemannian metric on $M$ of Lorentz signature $(1,3)$.
\footnote{The stated signature condition is equivalent to the requirement that, at every point $p$ in $M$, the tangent space $M_p$ has a basis \ $\overset{_1}{\xi} \ \! \!^a ,...,
\overset{_4}{\xi} \ \!\!^a$  \ such that, for all $i$ and $j$ in $\{1,2,3,4\}$,
\[
\overset{_i}{\xi} \ \! \!^a \ \overset{_i}{\xi} \ \! \!_a   = \left\{
\begin{array}{ll}
+1 & \mbox{if  \ $i=1$}\\
-1 & \mbox{if \  $i = 2, \, 3, \,  4$}
\end{array}
\right.
\]
and
$\overset{_i}{\xi} \ \! \!^a \ \overset{_j}{\xi} \ \! \!_a   = 0$ if  \ $i\ne j$. 
(Here we are using the abstract-index notation.  `$a$' is an abstract index, while `$i$' and `$j$' are normal counting indices.)   It follows that given any vectors  $\eta^a = \sum_{i=1}^4 \overset{_i}{k}  \ \overset{_i}{\xi} \ \! \!^a$, and
$\rho^a = \sum_{j=1}^4 \overset{_j}{l}  \ \overset{_j}{\xi} \ \! \!^a$ at $p$,
\begin{eqnarray*}
\eta^a \rho_a  & = &  
	\overset{_1}{k} \ \overset{_1}{l} - 	\overset{_2}{k} \ \overset{_2}{l} - 	\overset{_3}{k} \ \overset{_3}{l} - 	\overset{_4}{k} \ \overset{_4}{l}   \\
\eta^a \eta_a   & = &  \overset{_1}{k} \ \overset{_1}{k} - 	\overset{_2}{k} \ \overset{_2}{k} - 	\overset{_3}{k} \ \overset{_3}{k} - 	\overset{_4}{k} \ \overset{_4}{k}.   
\end{eqnarray*} 
\vspace{-1.5em}
}

We interpret $M$ as the manifold of point ``events" in the world.\footnote{We use `event' as a neutral term here and intend no special significance.   Some might prefer to speak of ``equivalence classes of coincident point events",
or  ``point event locations", or something along those lines.}  The interpretation of $g_{ab}$ is given by a network of interconnected physical principles. We list three in this section that are relatively simple in character because they make reference only to point particles and light rays. (These objects alone suffice to determine the metric, at least up to a constant.) In the next section, we list a fourth that concerns the behavior of (ideal) clocks. Still other principles involving generic matter fields will come up later.     

We begin by reviewing a few definitions.  In what follows, let $(M, g_{ab})$ be a fixed  relativistic spacetime, and let $\nabla_{\hspace{-.1em} a}$ be the derivative operator on $M$ determined by $g_{ab}$, i.e.,  the unique (torsion-free) derivative operator on $M$ satisfying the compatibility condition  $\nabla_{\! a} \, g_{bc} = \textbf{0}$.
 
 Given a point $p$ in $M$, and a vector $\eta^a$ in the tangent space $M_p$ at $p$,  we say $\eta^a$ is:
\[
\begin{array}{lllll}
     timelike  & \textrm{if} & \eta^a \eta_a  &  > & 0 \\
      null  \  (or \  lightlike) & \textrm{if} & \eta^a \eta_a  & = & 0  \\     
            causal  & \textrm{if} & \eta^a \eta_a  &  \geq & 0 \\ 
               spacelike  & \textrm{if} & \eta^a \eta_a  &  < & 0.         
    \end{array}
    \]
In this way, $g_{ab}$ determines a ``null-cone structure" in the tangent space at every point of $M$.  Null vectors form the boundary of the cone. Timelike vectors form its interior.  Spacelike vectors fall outside the cone.  Causal vectors are those that are either timelike or null. 
This classification extends naturally to \emph{curves}.  We take these to be smooth maps of the form $\gamma \! \! : I \rightarrow M$ where $I \subseteq \mathbb{R}$ is a (possibly infinite, not necessarily open) interval.\footnote{If $I$ is not an open set, we can understand smoothness to mean that there is an open interval $\overline{I} \subseteq \mathbb{R}$, with $I \subset \overline{I} $, and a smooth map $\overline{\gamma} \! \! : \overline{I} \rightarrow M$, such that $\overline{\gamma}(s) = \gamma(s)$ for all $s \in I$.}   $\gamma$ qualifies  as  \emph{timelike} (respectively \emph{null},  \emph{causal}, \emph{spacelike}) if its  tangent vector field $\vec{\gamma}$ is of this character at every point. 

A curve $\gamma_2 \! \! : I_2 \rightarrow M$ is called an \emph{(orientation preserving) reparametrization} of the curve $\gamma_1 \! \! : I_1 \rightarrow M$ if there is a smooth map $\tau \! \! : I_2 \rightarrow I_1$ of $I_2$ onto $I_1$, with positive derivative everywhere, such that $\gamma_2 = (\gamma_1 \circ \tau)$.  The property of being timelike, null, etc. is preserved under reparametrization.\footnote{This follows from the fact that, in the case just  described, $\vec{\gamma}_2  =  \frac{d\tau}{ds} \ \vec{\gamma}_1$, with $\frac{d\tau}{ds} > 0$. } So there is a clear sense in which our  classification also extends to \emph{images} of curves.\footnote{The difference between curves and curve images, i.e., between maps  $\gamma \! \! : I \rightarrow M$ and  sets $\gamma[I]$, matters. We take worldlines  to be instances of the latter, i.e., construe them as point sets rather than parametrized point sets.}  
   
A curve $\gamma \! \! : I \rightarrow M$  is said to be a \emph{geodesic (with respect to $g_{ab}$)} if its tangent field $\xi^a$ satisfies the condition:   $\xi^n  \nabla_{\! n} \, \xi^a = \textbf{0}$. 
The property of being a geodesic  is not, in general, preserved under reparametrization. So it does not transfer to curve images.  But, of course, the related property of being a \emph{geodesic up to reparametrization} does carry over.  (The latter holds of a curve if it can be reparametrized so as to be a geodesic.) 

Now we can state the first three interpretive principles. For all curves  $\gamma \!  : I \rightarrow M$, 
\begin{enumerate}
\item [C1]  $\gamma$ is timelike iff its image  $\gamma [I]$ could be  the worldline of a massive point particle (i.e., a particle with positive mass);\footnote{We will later discuss the concept of mass in relativity theory. For the moment, we take it to be just a primitive attribute of particles.}
\item[C2] $\gamma$ can be reparametrized so as to be a null geodesic iff $\gamma [I]$ could be the trajectory of a light ray;\footnote{For certain purposes, even within classical relativity theory, it is useful to think of light as constituted by streams of  ``photons", and take the right side condition here to be ``$\gamma [I]$ could be the worldline of a photon". The latter formulation makes C2 look more like C1 and P1, and draws attention to the fact that the distinction between massive particles and mass $0$ particles (such as photons) has direct significance in terms of relativistic spacetime structure. 
}

\item[P1] $\gamma$ can be reparametrized so as to be a timelike geodesic iff  $\gamma [I]$ could be  the worldline of a \emph{free}\footnote{``Free  particles" here must be understood as ones that do not experience any forces (except ``gravity").  It is one of the fundamental principles of relativity theory that gravity arises as a manifestation of  spacetime curvature, not as an external force that deflects particles from their natural, straight (geodesic) trajectories. We will discuss this matter further in section \ref{matter fields}.}  massive point particle.
\end{enumerate}  
\noindent  In each case, a statement about geometric structure (on the left) is correlated with a statement about the behavior of particles or light rays (on the right).

Several comments and qualifications are called for.  First,  we are here working within the framework of relativity as traditionally understood, and ignoring speculations about the possibility of particles (``tachyons") that travel faster than light.  (Their worldlines would come out as images of spacelike curves.)   Second, we have built in the requirement that ``curves" be smooth.  So, depending on how one models collisions of point particles, one might want to restrict attention here to particles that do not experience collisions. 

Third,  the assertions  require qualification because the status of ``point particles" in relativity theory is a delicate matter.  At issue is whether one treats a particle's own mass-energy as a source for the surrounding metric field $g_{ab}$  -- in addition to  other sources that may happen to be present.  (Here we anticipate our  discussion of Einstein's equation.) If one does, then the curvature associated with $g_{ab}$  may blow up as one approaches the particle's worldline.  And in this case one \emph{cannot} represent the worldline as the image of a curve in $M$, at least not without giving up the requirement that $g_{ab}$ be a smooth  field  on $M$. For this reason, a more careful formulation of the principles would restrict attention to ``test particles", i.e., ones whose own mass-energy is negligible and may be ignored for the purposes at hand.

Fourth, the modal character of the assertions (i.e., the reference to possibility)  is essential. It is simply not true, to take the case of C1,  that all timelike curve images \emph{are}, in fact, the worldlines of  massive particles. The claim is that, as least so far as the laws of relativity theory are concerned, they \emph{could} be.  Of course, judgments concerning what could be the  case depend on what conditions are held fixed in the background. 
The claim that a particular curve image could be the worldline of a massive point particle must be understood to mean that it could so long as there are, for example, no  barriers in the way.  Similarly, in C2  there is an implicit qualification. We are considering what trajectories are available to light rays when no intervening material media are present, i.e., when we are dealing with light rays \emph{in vacua}.

Though these four concerns are important and raise interesting questions about the role of idealization and modality in the formulation of physical theory,  they have little to do with relativity theory as such. Similar difficulties arise when one attempts to formulate corresponding principles within the framework of Newtonian gravitation theory.

It follows from the cited interpretive principles that the metric $g_{ab}$ is determined (up to a constant) by the behavior of point particles and  light rays.   We make this claim precise in a pair of propositions about ``conformal structure" and ``projective structure".

Let $\bar{g}_{ab}$ be a second smooth metric of Lorentz signature on $M$.   We say that $\bar{g}_{ab}$ is \emph{conformally equivalent} to $g_{ab}$ if there is a smooth map $\Omega \! : M \rightarrow \mathbb{R}$ on $M$ such that $\bar{g}_{ab} = \Omega^2 g_{ab}$.  ($\Omega$ is called a \emph{conformal factor}. It certainly need not be constant.) Clearly, if $\bar{g}_{ab}$ and $g_{ab}$ are conformally equivalent, then they agree in their classification of vectors and curves as timelike, null, etc..  The converse is true as well.\footnote{If the two metrics agree as to which vectors and curves belong to any one of the three categories, then they must agree on all. And in that case, they must be conformally equivalent. See Hawking and Ellis \shortcite[p.÷ 61]{Hawking-Ellis}.} 
Conformally equivalent metrics on $M$ do not agree, in general,  as to which curves on $M$ qualify as geodesics or even just as geodesics up to reparametrization.   But, it turns out, they do necessarily agree as to which \emph{null} curves are geodesics up to reparametrization.\footnote{This follows because the property of being the image of a null geodesic can be captured in terms of the existence or non-existence of (local) timelike and null curves connecting points in $M$. The relevant technical lemma can be formulated as follows. 
\begin{quote}
A curve $\gamma \! \! : I \rightarrow M$  can be reparametrized 
so as to be a null geodesic  iff $\gamma$ is null  and for all $s \in I$, there is an open set $O \subseteq M$ containing $\gamma(s)$ such that, for all $s_1, s_2 \in I$, if $s_1 \le s \le s_2$, and if $\gamma ([s_1, s_2 ]) \subseteq O$, then there is no timelike curve from $\gamma(s_1)$ to $\gamma(s_2)$  within $O$. 
\end{quote}
(Here $\gamma ([s_1, s_2 ])$ is the image of   $\gamma$ as restricted to the interval  $[s_1, s_2 ]$.)  For a proof, see Hawking and Ellis \shortcite[p.÷ 103]{Hawking-Ellis}.}  And the converse is true, once again.\footnote{For if the metrics agree as to which curves are null geodesics up to reparametrization, they must agree as to which vectors at arbitrary points are null, and this, we know, implies that the metrics are conformally equivalent.} 

Putting the pieces together, we have the following proposition. Clauses (1) and (2) correspond to C1 and C2 respectively.
\begin{proposition} \label{conformal structure} 
\ Let $\bar{g}_{ab}$ be a second smooth metric of Lorentz signature on $M$. Then the following conditions  are equivalent.  
\vspace{-.5em}
\begin{enumerate}
\vspace{-.5em}
\item [(1)]  $\bar{g}_{ab}$ and $g_{ab}$ agree as to which curves on $M$ are timelike.   
\vspace{-.5em}
\item [(2)]  $\bar{g}_{ab}$ and $g_{ab}$ agree as to which curves on $M$ can be reparameterized so as to be null geodesics. 
 \vspace{-.5em}
\item [(3)] $\bar{g}_{ab}$ and $g_{ab}$ are conformally equivalent.
\vspace{-.5em}
\end{enumerate}
\end{proposition}
\noindent In this sense,   \emph{the spacetime metric $g_{ab}$ is determined up to a conformal factor, independently, by the set of possible worldlines of massive point particles, and by the  set of possible trajectories of  light rays.}   

Next we turn to projective structure. Let $\overline{\nabla}_a$ be a second derivative operator on $M$.  We say that $\overline{\nabla}_a$ and $\nabla_a$ are \emph{projectively equivalent} if they agree as to which curves are geodesics up to reparametrization (i.e., if, for all curves $\gamma$, $\gamma$ can be reparametrized so as to be a geodesic with respect to $\overline{\nabla}_a$ iff it can be so reparametrized with respect to $\nabla_a$).  And if $\bar{g}_{ab}$  is a second metric on $M$ of Lorentz signature, we say that it is \emph{projectively equivalent} to $g_{ab}$ if its associated derivative operator $\overline{\nabla}_a$ is projectively equivalent to $\nabla_a$.  

It is a basic result, due to Hermann Weyl \shortcite{Weyl/article},
that if  $\bar{g}_{ab}$ and $g_{ab}$  are conformally \emph{and} projectively equivalent, then the conformal factor that relates them must be constant. It is convenient for our purposes, with interpretive principle P1 in mind, to cast it in a slightly altered form that makes reference only to timelike geodesics (rather than arbitrary geodesics).
\begin{proposition} \label{projective structure}  
Let $\bar{g}_{ab}$ be a second smooth metric on $M$ with $\bar{g}_{ab} = \Omega^2 g_{ab}$. If $\bar{g}_{ab}$ and $g_{ab}$ agree as to which timelike curves can be reparametrized so as to be geodesics, then $\Omega$ is constant.    
\end{proposition}
The spacetime metric $g_{ab}$, we saw, is determined  up to a conformal factor, independently, by the set of possible worldlines of massive point particles, and by the  set of possible trajectories of light rays.  The proposition now makes clear the sense in which  it is fully determined (up to a constant) by those sets together with the set of possible worldlines of free massive particles.\footnote{As Weyl put it  \shortcite[p.÷  103]{Weyl/PhilMath},
\begin {quote}
... it can be shown that the metrical structure of the world is already fully determined by its inertial and causal structure, that therefore mensuration need not depend on clocks and rigid bodies but that light signals and mass points moving under the influence of inertia alone will suffice.  
\end{quote}    
(For more on Weyl's  ``causal-inertial" method of determining the spacetime metric,  see Coleman and Kort\'e \shortcite[section 4.9]{Coleman-Korte}.)}  

Our characterization of relativistic spacetimes is extremely loose.  Many further conditions might be imposed. For  the moment, we consider just one.  

$(M, g_{ab})$ is said to be \emph{temporally orientable} if there exists a continuous timelike vector field $\tau^a$ on $M$.  Suppose the condition is satisfied. Then two such fields $\tau^a$ and $\hat{\tau}^a$ on $M$  are said to be \emph{co-oriented} if $\tau^a \hat{\tau}_a > 0$ everywhere, i.e., if $\tau^a$ and $\hat{\tau}^a$ fall in the same lobe of the null-cone at every point of $M$.  Co-orientation is an equivalence relation (on the set of continuous timelike vector fields on $M$) with two equivalence classes.  A \emph{temporal orientation} of $(M, g_{ab})$ is a choice of one of those two equivalence classes to count as the ``future" one. Thus, a non-zero causal  vector  $\xi^a$ at a point of $M$ is said to be \emph{future directed} or \emph{past directed} with respect to the temporal orientation $\mathcal{T}$ depending on whether  $\tau^a \xi_a > 0$ or $\tau^a \xi_a < 0$ at the point,  where $\tau^a$ is any continuous timelike vector field in $\mathcal{T}$. Derivatively, a causal curve  $\gamma \! \! : I \rightarrow M$ is said to be \emph{future directed} (resp.  \emph{past directed}) with respect to $\mathcal{T}$ if its tangent vectors at every point are.

In what follows, we assume that our background spacetime $(M, g_{ab})$ is temporally orientable, and that a particular temporal orientation has been specified. Also, given events $p$ and $q$ in $M$, we write \,  $p \ll q$ \, (resp. $p <q$)  if there is a future-directed timelike (resp. causal) curve that starts at $p$ and ends at $q$.\footnote{It follows immediately that if   $p \ll q$,    then   $p <q$. The converse does not hold, in general. But the only way the second condition can be true, without the  first being true as well, is if the only future-directed causal curves from $p$ to $q$ are null geodesics (or reparametrizations of null geodesics).  See Hawking and Ellis \shortcite[p.÷ 112]{Hawking-Ellis}.}    

\subsection{Proper Time}

So far we have discussed relativistic spacetime structure without reference to either ``time" or ``space". We come to them in this section and the next.  

Let $\gamma: [s_1, s_2] \rightarrow M$ be a future-directed timelike curve in $M$ with tangent field $\xi^a$. We associate with it an elapsed \emph{proper time}  (relative to $g_{ab}$) given by
\[
|\gamma| \,  =   \int_{s_1}^{s_2} (g_{ab} \,\xi^a  \, \xi^b)^\frac{1}{2} \ ds .
\] 
This elapsed proper time is invariant under reparametrization of $\gamma$, and is just what we would otherwise describe as the length of (the  image of) $\gamma$. The following is another basic principle of relativity theory. 
\begin{enumerate}
\item[P2] Clocks  record the passage of elapsed proper time along their worldlines.
\end{enumerate}

Again, a number of qualifications and comments are called for. Our formulation of C1, C2, and P1 was rough. The present formulation is that much more so.  We have taken for  granted that we know  what ``clocks" are. We have  assumed that they have worldlines (rather than worldtubes). And we have overlooked the fact that ordinary clocks (e.g., the  alarm clock on the  nightstand) do not do well at all when subjected  to extreme acceleration, tidal forces, and so forth. (Try smashing the alarm clock against the wall.)  Again, these concerns are important and raise interesting questions about the role of idealization in the formulation of physical theory. (One might construe an ``ideal clock" as a point-sized  test object that perfectly records the passage of proper time along its worldline, and then take P2 to assert that real clocks are, under appropriate conditions, to varying degrees of accuracy, approximately ideal.)  But as with our concerns about the status of point particles, they do not have much to do with relativity theory as such. Similar ones arise when one attempts to formulate corresponding principles about clock behavior within the  framework of Newtonian theory.

Now suppose that one has determined the conformal structure of spacetime, say, by using light rays. Then one can use clocks, rather than free particles, to determine the conformal factor.  One has the following simple result, which should  be compared with proposition \ref{projective structure}.\footnote{Here we not  only determine the metric up  to a constant, but determine the constant as well. 
The  difference is that here, in effect, we have built in a choice of units for spacetime distance.  We could obtain a more exact counterpart to proposition \ref{projective structure} if we worked, not with intervals of elapsed proper time,  but rather with ratios of such intervals. 
}

\begin{proposition} \label{clocks} 
Let $\bar{g}_{ab}$ be a second smooth metric on $M$ with $\bar{g}_{ab} = \Omega^2 g_{ab}$. Further suppose that the two metrics assign the same lengths to all timelike curves, i.e.,  $|\gamma|_{\bar{g}_{ab}} =  |\gamma|_{g_{ab}}$ for all timelike curves $\gamma: I \rightarrow M $. Then $\Omega = 1 $ everywhere.
(Here $|\gamma|_{g_{ab}}$  is the length of $\gamma$ relative to $g_{ab}$.)
\end{proposition}


P2    gives  the whole story of relativistic clock behavior (modulo the concerns noted above).  In particular, it implies the path dependence of clock readings.   
If two clocks start at an event $p$, and travel along different trajectories to an event $q$, then, in general, they will record different elapsed times for the trip. (E.g., one will record an elapsed time of 3,806 seconds, the other 649 seconds.) This is true no matter how similar the clocks are. (We may stipulate that they came off the same assembly line.)  This is the case because, as P2 asserts,  the elapsed time recorded by each of the clocks is  just the length of the timelike curve it traverses in getting from $p$ to $q$ and, in general, those lengths will be different.
	
Suppose we consider all future-directed timelike curves from $p$ to $q$. It is natural to ask if there are any that minimize or maximize the recorded elapsed time between the events. The answer to the first question is `no'. Indeed, one has the following proposition.

\begin{proposition} \label{no minimal time} 
Let $p$ and $q$ be events in $M$ such that  $p \ll q$. Then, for all $\epsilon > 0$, there exists a future-directed timelike curve $\gamma$ from $p$ to $q$ with $| \gamma | < \epsilon$. (But there is no such curve with length $0$, since all timelike curves have non-zero  length.)
\end{proposition}

Though some work is  required to give the proposition an honest proof (see O'Neill \shortcite[pp.÷  294-5]{O'Neill}), it should seem intuitively plausible. If there is a timelike curve connecting $p$ and $q$, there also exists a jointed, zig-zag null curve that connects them. 
\begin{figure}[h]
\begin{center}
\setlength{\unitlength}{1cm}
\begin{picture}(1.85,4.3)
 \put(0,0.0){\epsfig{figure=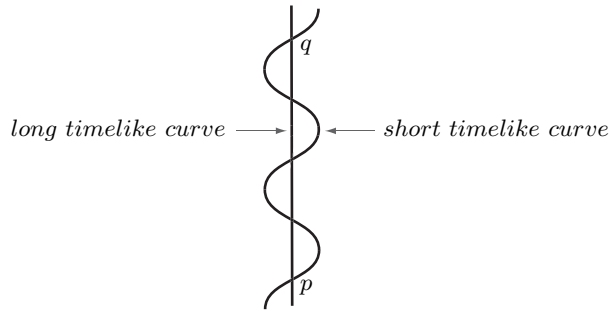}}
\put(.85,.25){\small $p$}
\put(.85,3.45){\small $q$}
\put(1.95,2.30){\small $short \ timelike \  curve$}
\put(-3.0,2.30){\small $long \ timelike \  curve$}
\end{picture} 

\vspace{-1em}
\begin{minipage}[t]{9.0cm}
\renewcommand{\baselinestretch}{1.0}
\caption{A long timelike curve from $p$ to $q$ and a very short one  that swings back-and-forth, and approximates a broken null curve.}\label{wigglecurve}
\vspace{-.5em}
\end{minipage}
\end{center}
\end{figure}
 It has length $0$. But we can approximate the jointed null curve arbitrarily closely with smooth timelike curves that swing back and forth.  So (by the continuity of the length function), we should  expect that, for all $\epsilon > 0$, there is an approximating timelike curve that has length less than $\epsilon$.  (See figure \ref{wigglecurve}.)  
 
The answer to the second question (Can one \emph{maximize} recorded elapsed time between $p$ and $q$?) is `yes' if one restricts attention to local regions of spacetime. In the case of positive definite metrics, i.e., ones with signature of form $(n,0)$, we know, geodesics are \emph{locally shortest} curves. The corresponding result for Lorentz metrics is that timelike geodesics are \emph{locally longest}  curves.

\begin{proposition} \label{local maximal time} 
Let $\gamma : I \rightarrow M$  be a future-directed timelike curve.  Then $\gamma$ can be reparametrized so as to be a geodesic iff for all $s \in I$, there exists an open set $O$ containing $\gamma(s)$ such that, for all $s_1, s_2 \in I$ with $s_1 \le s \le s_2$, if the image of $\overline{\gamma} = \gamma _{_{| [s_1, s_2]}}$ is contained in  $O$, then $\overline{\gamma}$ (and its reparametrizations) are longer than all other timelike curves in $O$ from $\gamma(s_1)$ to $\gamma(s_2)$.  $($Here  $\gamma _{_{| [s_1, s_2]}}$   is the restriction of $\gamma$ to the interval  $ [s_1, s_2]$.$)$
\end{proposition}
The proof of the proposition is very much the same as in the positive definite case. (See Hawking and Ellis \shortcite[p.÷  105]{Hawking-Ellis}.) Thus of all clocks passing locally from $p$ to $q$, that one will record the greatest elapsed time that ``falls freely" from $p$ to $q$. To get a clock to read a smaller elapsed time than the maximal value one will have to accelerate the clock.  Now acceleration requires fuel, and fuel is not free. So proposition \ref{local maximal time} has the consequence that (locally) ``saving time costs money". And  proposition \ref{no minimal time} may be taken to imply that (locally) ``with enough money one can save as much time as one wants".


The restriction here to local regions of spacetime is essential. The connection described between clock behavior and acceleration does not, in general, hold on a global scale.  In some relativistic spacetimes, one can find future-directed  timelike geodesics connecting two events that have different  lengths, and so clocks following the curves will record different elapsed times between the events even though \emph{both} are in a state of free fall. 
Furthermore -- this follows from the preceding claim by continuity considerations alone -- it can be the case that of two clocks passing between the events, the one that undergoes acceleration during the trip records a \emph{greater} elapsed time than the one that remains in a state of free fall.

The connection we have been considering between clock behavior and acceleration was once thought to be paradoxical.  (I am thinking of the ``clock (or twin) paradox".)  Suppose two clocks, A and B, pass from one event to another in a suitably small region of spacetime. Further suppose A does so in a state of free fall, but B undergoes acceleration at some point along the way. Then, we know, A will record a greater elapsed time for the trip than B.  This was thought paradoxical because it was believed that ``relativity theory denies the possibility of distinguishing ``absolutely" between free fall motion and accelerated motion".  (If we are equally well entitled to think that it is clock  B that is in a state of free fall, and A  that undergoes acceleration, then, by parity of reasoning, it should be B that records the greater elapsed time.) The resolution of the paradox, if one can call it that, is that relativity theory makes no such denial. The situations of A and B here are \emph{not} symmetric. The distinction between accelerated motion and free fall makes every bit as much sense in relativity theory as it does in Newtonian physics. 
   

In what follows, unless indication is given  to the  contrary,  a ``timelike curve" should be understood to be a future-directed timelike curve, parametrized by elapsed proper time, i.e., by arc length.  In that case, the tangent field $\xi^a$ of the curve has unit length ($\xi^a \xi_a = 1$).  And if a particle happens to have the image of the curve as its worldline, then, at any point, $\xi^a$ is called the  particle's \emph{four-velocity}  there.   

\subsection{Space/Time Decomposition at a Point and Particle Dynamics} \label{Sp-Time Decomp}

Let $\gamma$ be a timelike curve representing the particle $O$ with four-velocity field $\xi^a$.   Let $p$ be a point on the image of $\gamma$, and let $\lambda^a$ be a vector at $p$.   There is a natural decomposition of $\lambda^a$ into components parallel to, and orthogonal to, $\xi^a$:

\begin{equation}
\label{vectordecomposition}
\lambda^a  =   \underbrace{(\lambda^b \xi_b) \xi^a}_{parallel \ to \ \xi^a}   +  \  \underbrace{(\lambda^a - (\lambda^b \xi_b) \xi^a)}_{orthogonal \ to \ \xi^a} . 
\end{equation}
 
\noindent These are standardly interpreted, respectively, as the ``temporal" and ``spatial" components of  $\lambda^a$ (relative to $\xi^a$).  In particular, the three-dimensional subspace of $M_p$ consisting of  vectors orthogonal to $\xi^a$  is interpreted as the ``infinitesimal" simultaneity slice of $O$ at $p$.\footnote{Here we simply take for granted the standard identification  of  ``relative simultaneity" with orthogonality. We will return to consider its justification in section \ref{simultaneity}. }   
%
%
\noindent If we introduce the tangent and orthogonal projection operators
\vspace{-1em}
\begin{eqnarray}
k_{ab} & = &  \xi_a \, \xi_b     \label{def k}\\
h_{ab} & = & g_{ab} -  \xi_a  \, \xi_b   \label{def h} 
\end{eqnarray}
then the decomposition can be  expressed in the form
\begin{equation}
\label{}
\lambda^a  =      k^a_{\ b} \, \lambda^b   +  h^a_{\ b}  \, \lambda^b.  \label{second vector decomposition}
\end{equation}
We can think of $k_{ab}$ and $h_{ab}$ as the relative temporal and spatial metrics determined by $\xi^a$. They are symmetric and satisfy
\begin{eqnarray}
k^a_{\ b} \, k^b_{\ c} & = &  k^a_{\ c}     \label{multiple k}\\ 
h^a_{\ b} \, h^b_{\ c} & = &  h^a_{\ c}.  \label{multiple h}
\end{eqnarray}

Many standard textbook assertions concerning the kinematics and dynamics of point particles can be recovered using these decomposition formulas. For example, suppose that the worldline of a second particle $\overline{O}$ also passes through $p$ and that its four-velocity at $p$ is $\overline{\xi}\hspace{.05em}^{a}$.  (Since $\xi^a$ and $\overline{\xi}\hspace{.05em}^{a}$ are both future-directed, they are co-oriented, i.e., $(\xi^a \, \overline{\xi}\hspace{.05em}_{a}) > 0$.) We compute the speed of $\overline{O}$ as determined by $O$. To do so, we take the spatial magnitude of $\overline{\xi}\hspace{.05em}^a$  relative to $O$ and divide by its temporal magnitude relative to $O$:
\begin{equation}
\label{v-def }
v = speed \ of  \  \overline{O} \   relative \   to \   O = \frac{\|h^a_{\ b} \,\overline{\xi}\hspace{.05em}^b \|}{\|k^a_{\ b} \,\overline{\xi}\hspace{.05em}^b \|}. 
\end{equation}
(Given any vector $\mu^a$,  we understand  $\|\mu^a\|$ to be
  $(\mu^a \mu_a)^\frac{1}{2}$ if $\mu^a$ is causal, and  $(-\mu^a \mu_a)^\frac{1}{2}$ if it is spacelike.)  
From (\ref{def k}), (\ref{def h}), (\ref{multiple k}), and (\ref{multiple h}), we have
\begin{equation}
\label{k-component}
\|k^a_{\ b} \,\overline{\xi}\hspace{.05em}^b \|= (k^a_{\ b} \, \overline{\xi}\hspace{.05em}^b \, k_{ac} \, \overline{\xi}\hspace{.05em}^c)^\frac{1}{2} = (k_{bc} \, \overline{\xi}\hspace{.05em}^b \, \overline{\xi}\hspace{.05em}^c)^\frac{1}{2} = (\overline{\xi}\hspace{.05em}^b \, \xi_b)   
\end{equation}
and
\begin{equation}
\label{h-component}
\|h^a_{\ b} \,\overline{\xi}\hspace{.05em}^b \|= (-h^a_{\ b} \, \overline{\xi}\hspace{.05em}^b \, h_{ac} \, \overline{\xi}\hspace{.05em}^c)^\frac{1}{2} = (-h_{bc} \, \overline{\xi}\hspace{.05em}^b \, \overline{\xi}\hspace{.05em}^c)^\frac{1}{2} = ((\overline{\xi}\hspace{.05em}^b \, \xi_b)^2 - 1)^{\frac{1}{2}}.  
\end{equation} 
So 
\begin{equation}
\label{v=}
v = \frac{((\overline{\xi}\hspace{.05em}^b \, \xi_b)^2 - 1)^{\frac{1}{2}}}
{(\overline{\xi}\hspace{.05em}^b \, \xi_b)}   <  1 . 
\end{equation}
Thus, as measured by $O$, no massive particle can ever attain the maximal speed 1.   (A similar calculation would show that, as determined by $O$,  light always travels with speed 1.) For future reference, we note that (\ref{v=}) implies:
\begin{equation}
\label{Inversion of v =}
\overline{\xi}\hspace{.05em}^b \, \xi_b  = \frac{1}{\sqrt{1 \,  - \, v^2}}.  
\end{equation} 
\begin{sloppypar}
It is a basic fact of relativistic life that there is associated with every point particle, at every event on its worldline, a \emph{four-momentum} (or \emph{energy-momentum}) vector $P^a$. In the case of a massive particle with four-velocity $\overline{\xi}\hspace{.05em}^ a$,  $P^a$ is proportional to  $\overline{\xi}\hspace{.05em}^a$,  and the (positive) proportionality factor  is just what we would otherwise call the \emph{mass} (or \emph{rest mass}) $m$ of the particle. So we have   $P^a = m \ \overline{\xi}\hspace{.05em}^a$.   In the case of a ``photon" (or other mass $0$ particle),  no such characterization is available because its worldline is the image of a null (rather than timelike) curve.
But we can still understand its four-momentum vector at the event in question to be a future-directed null vector that is tangent to its worldline there.  If we think of the four-momentum vector $P^a$ as fundamental, then we can, in both cases, recover the mass of the  particle as the length of $P^a$:  $m =  (P^a P_a)^{\frac{1}{2}}$.  (It is strictly positive in the first case, and $0$ in the second.) 
\end{sloppypar} 

Now suppose a massive particle $O$ has four-velocity $\xi^a$ at an event, and another particle, either a massive particle or a photon,  has four-momentum $P^a$ there. We can recover the usual expressions for the energy and three-momentum of the second particle relative to $O$ if we decompose $P^a$ in terms of $\xi^a$. By (\ref{second vector decomposition}) and (\ref{def k}),  we have 
\begin{equation}
\label{Pdecomposition}
P^a  =   (P^b \xi_b) \, \xi^a  \ +  \ h^a_{\ b} P^b.
\end{equation}
The \emph{energy} relative to $O$ is the coefficient in the first term:  $E =  P^b \xi_b$.  In the case of a massive particle where  $P^a = m \, \overline{\xi}\hspace{.05em}^a$, this yields, by  (\ref{Inversion of v =}),   

\begin{equation}
\label{relative energy}
E = m \, (\overline{\xi}\hspace{.05em}^b \, \xi_b)   =  \frac{m}{\sqrt{1 \,  - \, v^2}}.
\end{equation}  

\noindent (If we had not chosen units in which $c = 1$, the numerator in the final expression would have been $m c^2$ and the denominator $\sqrt{1   -  \frac{v^2}{c^2}}$.)  The \emph{three-momentum} relative to $O$ is the second term in the decomposition, i.e., the component of $P^a$ orthogonal to $\xi^a$:    $h^a_{\ b} \,  P^b$. In the case of a massive particle, by (\ref{h-component}) and  (\ref{Inversion of v =}),  it has magnitude

\begin{equation}
\label{relative momentum}
p  = \|h^a_{\ b} \,  m \, \overline{\xi}\hspace{.05em}^b\| =  m \, ((\overline{\xi}\hspace{.05em}^b \, \xi_b)^2 - 1)^{\frac{1}{2}} = \frac{m \, v}{\sqrt{1 \,  - \, v^2}}. 
\end{equation} 

Interpretive principle P1 asserts that free particles traverse the images of timelike geodesics. It can be thought of as the relativistic version of Newton's first law of motion. Now we consider acceleration and the relativistic version  of the second law.   Let $\gamma : I \rightarrow M$ be a  timelike curve whose image is the worldline of a massive particle $O$,  and let $\xi^a$ be the four-velocity field of $O$. Then the \emph{four-acceleration} (or just $\emph{acceleration}$) field of $O$ is $\xi^n  \nabla_n \, \xi^a$, i.e.,  the directional derivative of $\xi^a$ in the direction $\xi^a$. The four-acceleration vector is orthogonal to $\xi^a$. (This is clear, since $\xi^a (\xi^n  \nabla_n \, \xi_a)  \, = \, \frac{1}{2} \, \xi^n  \nabla_n \, (\xi^a \, \xi_a)  \, = \, \frac{1}{2} \, \xi^n  \nabla_n \, ( 1)  \, = \,  0.$)  The magnitude \ $\|\xi^n  \nabla_n \, \xi^a \|$ \  of the four-acceleration vector at a point   is just what we would otherwise describe as the Gaussian curvature of $\gamma$ there.   It is a measure of the degree to which $\gamma$ curves away from a straight  path.  (And $\gamma$ is a geodesic precisely if its curvature vanishes everywhere.) 
%
%

The notion of spacetime acceleration requires attention. Consider an example. Suppose you decide to end it all and jump  off the Empire State Building.  What would your acceleration  history be like during your  final moments?  One is  accustomed in such cases to think in terms of acceleration relative to the earth.  So one would  say that you undergo acceleration between the time of your jump and your calamitous arrival. But on the present account, that description has things backwards.  Between jump and arrival you are \emph{not} accelerating. You are in a state of free fall and moving (approximately) along a spacetime geodesic. But before the jump, and after the arrival, you \emph{are} accelerating. The floor of the observation desk, and then later the sidewalk, push you away from a geodesic  path.  The all-important idea here is  that we are incorporating the ``gravitational field" into the geometric structure of spacetime, and particles traverse geodesics if and only if they are acted upon by no forces ``except gravity".

The acceleration of any massive particle, i.e., its deviation from a geodesic trajectory,  is determined by the forces acting on it (other than ``gravity").  If the particle has mass $m > 0$, and the vector field $F^a$ on $\gamma[I]$ represents the vector sum of the various (non-gravitational) forces acting on the particle, then the particle's four-acceleration $\xi^n \,\nabla_n \, \xi^a$ satisfies: 
\begin{equation} \label{Second Law}
F^a = m \ \xi^n \nabla_n \, \xi^a.
\end{equation} 
\noindent This is our version of Newton's second law of  motion. 

Consider an example.  Electromagnetic fields are represented by smooth, anti-symmetric fields $F_{ab}$. (Here ``anti-symmetry" is the condition that $F_{ba} = - F_{ab}$.)   If a particle with mass $m > 0$, charge $q$, and four-velocity field $\xi^a$ is present,  the force exerted by the field on the particle at a point is given by  $q \, F^{a}_ {\ b} \, \xi^b$.  If we use this expression for the left side of (\ref{Second Law}), we arrive at  the Lorentz law of motion for charged particles in the presence of an electromagnetic field:   
\begin{equation}  \label{Lorentz.equation.motion}
q \, F^{a}_ {\ b} \, \xi^b \, = \,  m \ \xi^b \, \nabla_b \, \xi^a.\footnote{Notice that the equation makes geometric sense. The acceleration vector on the right is orthogonal to  $\xi^a$. But so is the force vector on the left since  $\xi_a  ( F^{a}_ {\ b} \, \xi^b) = \xi^a \, \xi^b  F_{ab} =  \frac{1}{2} \,  \xi^a  \xi^b \,  (F_{ab} + F_{ba})$, and by the anti-symmetry of $F_{ab}$,  $(F_{ab} + F_{ba}) = \mathbf{0}.$}   
\end{equation}

\subsection{Matter Fields}\label{matter fields}

In classical relativity theory, one generally takes for granted that all that there is, and all that happens, can be described in terms of various \emph{matter fields}, e.g., material fluids and electromagnetic fields.\footnote{This being the case, the question arises how (or whether) one can adequately recover talk about ``point particles" in terms of the matter fields. We will say just a bit about the question in this section.}  Each such field is represented by one or more smooth tensor (or  spinor) fields on the spacetime manifold $M$. Each is assumed to satisfy field equations involving the fields that represent it and the spacetime metric $g_{ab}$.  

For present purposes, the most important basic assumption about the matter fields is the following. 
\begin{quotation} 
\noindent Associated with each matter field $\mathcal{F}$  is a symmetric smooth tensor field $T_{ab}$ characterized by the property that, for all points $p$ in $M$, and all future-directed, unit timelike vectors $\xi^a$ at $p$,  $T^a_{\ \  b} \,  \xi^b$  is the four-momentum density of $\mathcal{F}$ at $p$ as determined relative to $\xi^a$.
 \end{quotation}   
\noindent  $T_{ab}$ is called the \emph{energy-momentum} field associated with $\mathcal{F}$. The four-momentum density vector  $T^a_{\ \  b} \,  \xi^b$ at $p$ can be further decomposed into its temporal and spatial components relative to $\xi^a$, just as the four-momentum of a massive particle was decomposed in the preceding section.  The coefficient of $\xi^a$ in the first component,  $T_{ab} \, \xi^a  \xi^b$, is the \emph{energy density} of  $\mathcal{F}$ at $p$ as determined relative to $\xi^a$.  The second component, 
$T_{nb} \, (g^{an} - \xi^a \, \xi^n) \, \xi^b$, is the \emph{three-momentum density} of  $\mathcal{F}$ at $p$ as determined relative to $\xi^a$.  

Other assumptions about matter fields can be captured as constraints on the energy-momentum tensor fields with which they are associated. Examples are the following. (Suppose $T_{ab}$ is associated with matter field $\mathcal{F}$.)
\vspace{-.5em}
\begin{description}
\item [Weak Energy Condition:]  Given any future-directed unit timelike vector $\xi^a$ at any point in $M$,  $T_{ab} \, \xi^a \xi^b \geq 0$. 
\item[Dominant Energy Condition:] Given any future-directed unit timelike vector $\xi^a$ at any point in $M$,  $T_{ab}\,  \xi^a \xi^b \geq 0$ \emph{and} $T^a_{\ b} \, \xi^b $  is timelike or null.    
\item[Conservation Condition:]  $\nabla_a \, T^{ab} = \mathbf{0}$ at all points in M. 
\end{description} 
\noindent The first asserts that the energy density of $\mathcal{F}$, as determined by any observer at any point, is non-negative. The second adds the requirement that the four-momentum density of  $\mathcal{F}$, as determined by any observer at any point, is a future-directed causal (i.e., timelike or  null) vector.  It captures the condition that there is an upper bound to the speed with which energy-momentum can propagate (as determined by any observer).  It captures something of the flavor of principle C1 in section \ref{Relativistic  Spacetimes}, but avoids reference to ``point particles".\footnote{This is the standard formulation of the dominant energy condition. The fit with C1 would be even closer if we strengthened the condition slightly so as to be appropriate, specifically, for massive  matter fields:  at any point $p$ in $M$, if  $T^a_{\ b}  \neq \textbf{0} $ there, then  $T^a_{\ b} \, \xi^b $  is \emph{timelike} for all  future-directed unit timelike vectors $\xi^a$ at $p$.}

The conservation condition, finally,  asserts that the energy-momentum carried by $\mathcal{F}$ is locally conserved.  If two  or more matter fields are present in the same region of spacetime, it need not be the case that each one individually satisfies the condition. Interaction may occur.  But it \emph{is} a fundamental assumption that the composite energy-momentum field formed by taking the sum of the individual ones satisfies it.  Energy-momentum can  be transferred from one matter field to another, but it cannot  be created or destroyed.  

The dominant energy and conservation conditions have a number of joint consequences that support the interpretations just given. We mention two. The first requires a preliminary definition.  

	Let $(M, \,   g_{ab})$  be a  fixed relativistic spacetime, and let $S$ be an achronal subset of $M$ (i.e., a subset in which there do \emph{not} exist points $p$ and $q$ such that $p\ll q$).
The \emph{domain of dependence} $D(S)$ of $S$ is the set of all points $p$ in $M$ with this property: given any smooth causal curve without (past or future) endpoint,\footnote{Let  $\gamma:I \rightarrow  M$ be a smooth curve. We say that a point $p$ in $M$ is a \emph{future-endpoint} of $\gamma$ if, for all open sets $O$ containing $p$, there exists an $s_0$ in $I$ such that for all $s \in I$, if $s \ge s_0$, then $\gamma(s) \in O$, i.e.,  the image of $\gamma$ eventually enters and remains in $O$. (\emph{Past-endpoints} are defined similarly.)} if (its image) passes through $p$, then it necessarily intersects $S$. %
	%
%
\noindent  For all standard matter fields, at least, one can prove a theorem to the effect that ``what happens on $S$ fully determines what happens throughout $D(S)$". (See Earman (this volume, chapter 15).) Here we consider just a special case.

\begin{proposition} \label{energy-momentum/domain of dependence} 
Let $S$ be an achronal subset of $M$. Further let $T_{ab}$ be a smooth symmetric field on $M$ that satisfies both the dominant energy and conservation conditions. Finally, assume $T_{ab} = \mathbf{0}$ on $S$. Then $T_{ab} = \mathbf{0}$ on all of $D(S)$.
\end{proposition}

The intended interpretation of the proposition is clear.  If energy-momentum cannot propagate (locally) outside the null-cone, and if it is conserved, and if it vanishes on $S$, then it must vanish throughout $D(S)$. After all, how could it ``get to" any point in $D(S)$? Note that our formulation of the proposition does not presuppose any particular physical interpretation of the symmetric field $T_{ab}$. All that is required is that it satisfy the two stated conditions. (For a proof, see Hawking and Ellis \shortcite[p.÷   94]{Hawking-Ellis}.)

The next proposition (Geroch and Jang \shortcite{Geroch-Jang})  shows that, in a sense, if  one assumes the dominant energy condition and the conservation condition, then one can \emph{prove} that free massive point particles traverse the images of timelike geodesics. (Recall principle P1 in section \ref{Sp-Time Decomp}.)  The trick is to find a way to talk about ``point particles" in the language of extended matter fields.

\begin{proposition}  \label{Geroch-Jang} 
Let $\gamma: I \rightarrow M$ be smooth curve. Suppose that given any open subset $O$ of $M$ containing $\gamma [I]$, there exists a smooth symmetric field $T_{ab}$ on $M$ such that:
\begin{enumerate}
\vspace{-.8em}
\item[(1)]  $T_{ab}$ satisfies the dominant energy condition; 
\vspace{-.4em}
\item[(2)]  $T_{ab}$ satisfies the conservation condition;
\vspace{-.4em}
\item[(3)]  $T_{ab} = \mathbf{0}$ outside of $O$; 
\vspace{-.4em}
\item[(4)]  $T_{ab} \neq \mathbf{0}$ at some point in $O$.  
\end{enumerate}
\vspace{-.8em}
Then $\gamma$ is timelike, and can be reparametrized so as to be a geodesic.
\end{proposition}


The proposition might be paraphrased this way.  If a smooth curve in spacetime is such that arbitrarily small free bodies could contain the image of the curve in their worldtubes, then the  curve must be a timelike geodesic (up to reparametrization). In effect, we are trading in ``point particles" in  favor of nested convergent sequences of smaller and smaller extended particles. (Bodies here are understood to be ``free" if their internal energy-momentum is conserved. If a  body is acted upon by a field, it is only the composite energy-momentum of the body and field together that is conserved.) 

Note that our formulation of the proposition takes for granted that we can keep the background spacetime structure $(M, g_{ab})$ fixed while altering the fields $T_{ab}$ that live on $M$.   This is justifiable only to the extent that, in each case, $T_{ab}$ is understood to represent a test body whose effect on the background spacetime structure is negligible.\footnote{Stronger theorems have been proved (see Ehlers and Geroch \shortcite{Ehlers-Geroch}) in which it is not required that the perturbative effect  of the extended body disappear entirely at each stage of the limiting process, but only that,  in a certain sense,  it disappear in the limit.} 
Note also that we do  not have to assume at the outset that the curve $\gamma$ is timelike.  That follows from the other assumptions.

We have here a precise proposition in the language of matter fields that, at least to some degree, captures principle P1   (concerning the behavior of free massive point particles). Similarly, it is possible to capture C2 (concerning the behavior of  light) with a proposition about the behavior of solutions to Maxwell's equations in a limiting regime (``the geometrical limit") where wavelengths are small.  It asserts, in effect, that when one passes to this limit, packets of electromagnetic waves are constrained to move along (images of) null geodesics. (See Wald \shortcite [p.÷   71]{Wald}.) 

Now we consider an example.  \emph{Perfect fluids} are represented by three objects: a four-velocity field $\eta^a$, an energy density field $\rho$, and an isotropic pressure field $p$ (the  latter two as determined by a ``co-moving" observer at rest in the fluid).    In the special case where the pressure $p$ vanishes, one speaks of a \emph{dust field}.  Particular instances of  perfect fluids are characterized by ``equations of state"  that specify $p$ as a function of $\rho$. (Specifically excluded here are such complicating factors as anisotropic pressure, shear stress, and viscosity.)  Though $\rho$ is generally assumed to be non-negative (see below), some perfect fluids (e.g., to a good approximation, water)  can exert negative pressure. The energy-momentum tensor field associated with a perfect fluid is:
\begin{equation} \label{T-perfect fluid}
T_{ab} = \rho \, \eta_a \, \eta_b  -  p \, (g_{ab} \, - \, \eta_a \, \eta_b).  \end{equation} 
Notice that the energy-momentum density vector of the fluid at any point, as determined by a co-moving observer (i.e., as determined relative to $\eta^a$), is  $T^a_{\ b}  \, \eta^b = \rho \, \eta^a$. So we can understand  $\rho$, equivalently, as the energy density of the fluid relative to $\eta^a$, i.e., $T_{a b}  \, \eta^a \, \eta^b$,   or as the (rest) mass density of the fluid, i.e.,  the length of $\rho \, \eta^a$.  (Of course, the situation here corresponds to that of a  point particle with mass $m$ and four-velocity $\eta^a$, as considered in section \ref{Sp-Time Decomp}.)     

In the case of a perfect fluid, the weak energy condition (WEC), dominant energy condition (DEC), and conservation condition (CC) come out as follows.



\begin{center}
$
\begin{array}{ccl }
	\textrm{WEC} & \Longleftrightarrow & \rho \geq 0  \hspace{1em}  \textrm{and}   \hspace{3.2em}   p  \, \geq  \,  - \rho  \\ 
      \textrm{DEC} & \Longleftrightarrow & \rho \geq 0  \hspace{1em}  \textrm{and}   \hspace{1em}   \rho  \, \geq  \, p  \, \geq \,  -\rho \\

		{} & {}  & {}  \\ 
	\textrm{CC} & \Longleftrightarrow & \left\{
	\begin{array}{lcl}
	(\rho + p) \, \eta^b \, \nabla_b  \, \eta^a  - (g^{ab}  - \eta^a \, \eta^b) \, \nabla_b  \, p  &=& \mathbf{0}     \\
	\eta^b \, \nabla_b  \, \rho  \, + \,  (\rho + p)  \, (\nabla_b  \, \eta^b) &=&  0    
	\end{array}
	\right.  
\end{array}
$
\end{center}


Consider the two equations jointly equivalent to the conservation condition.  The first is the equation of motion for a perfect fluid. We can think of it as a relativistic version of Euler's equation.  The second is an equation of continuity (or conservation) in the sense familiar from classical fluid mechanics.  It is easiest to think about the special case of a dust field ($p = 0$). In that case, the equation of motion reduces to the geodesic equation: $\eta^b \, \nabla_b  \, \eta^a  = \textbf{0}$. That makes sense. In the absence of pressure, particles in the fluid are free particles.    And the conservation  equation reduces to: 
$\eta^b \, \nabla_b  \, \rho  \, + \,  \rho  \, (\nabla_b  \, \eta^b) = 0$. The first term gives the instantaneous rate of change of the fluid's energy density, as determined by a co-moving observer. The term $\nabla_b  \, \eta^b$ gives the instantaneous rate of change of its volume, per unit volume,  as determined by that observer.   In a more familiar notation, the equation might be written \ 
$\displaystyle\frac{d\rho}{ds} + \frac{\rho}{V} \frac{dV}{ds} = 0$ \  or, equivalently,  \ $\displaystyle \frac{d(\rho V)}{ds} = 0$. \ (Here we use $s$ for elapsed proper time.) It asserts that (in the absence of pressure, as determined by a co-moving observer)  the energy contained in an (infinitesimal)  fluid blob remains constant, even as its volume changes.  
  
In the general case, the situation is more complex because the pressure in the fluid contributes to its energy (as determined relative to particular observers), and hence to what might be called its ``effective mass density". (If you compress a fluid blob, it gets heavier.)  In this case, the WEC comes out as the requirement that $(\rho + p) \geq 0$ in addition to $\rho \geq 0$. If we take $h^{ab} = (g^{ab} - \eta^a \, \eta^b)$, the equation of motion  can be expressed as:  
\begin{equation*}
(\rho + p) \, \eta^b \, \nabla_b  \, \eta^a  \, = \,   h^{ab}  \, \nabla_b  \, p.
\end{equation*}
This is an instance of the ``second law of motion" (\ref{Second Law}) as applied to an (infinitesimal) blob of fluid.  On the left we have:   ``effective mass density  $\times$   acceleration". On the right, we have the force acting on the blob. We can think of it as minus\footnote{The minus sign comes in because of our sign conventions.} the gradient of the pressure (as determined by a co-moving observer).  Again, this makes sense. If the pressure on the left side of the blob is greater than that on the right, it will move to the right. The presence of  the non-vanishing term  $(p \, \nabla_b  \eta^b)$ in the conservation equation is now required because the energy of the blob is \emph{not} constant when its volume changes as a result of the pressure.  The equation governs the contribution made to its energy by pressure.

\subsection{Einstein's Equation} \label{Einstein's Equation}

Once again, let $(M, \,   g_{ab})$  be our background relativistic spacetime with a specified temporal orientation. 

It is  one of the fundamental ideas of relativity theory that spacetime structure is not a fixed backdrop against which the processes of physics unfold, but instead participates in that unfolding.   It posits a dynamical interaction between the spacetime metric in any region and the matter fields there. The interaction is governed by \emph{Einstein's field equation}:
\begin{equation}\label{EE regular}
R_{ab} \,  - \frac{1}{2} \, R \, g_{ab}   - \lambda \, g_{ab} \,  = \,  8 \, \pi \, T_{ab},  
\end{equation} 
\noindent or, equivalently,
\begin{equation}\label{EE reversed}
R_{ab} \, = \,  8 \, \pi \, (T_{ab}   - \frac{1}{2} \, T \, g_{ab})  \, - \,  \lambda \, g_{ab}.  
\end{equation} 
Here  $\lambda$ is the  \emph{cosmological constant}, $R_{ab} \,  (= R^n_{\ abn})$ is the \emph{Ricci tensor field}, $R \, (= R^a_{\ a})$ is the Riemann scalar curvature field, and $T$ is the contracted field $T^a_{\ a}$.\footnote{We use ``geometrical units" in which the gravitational constant $G$, as well as the speed of light $c$, is $1$.  
} We start with four remarks about (\ref{EE regular}), and then consider an alternative formulation that provides a geometric interpretation of sorts. 

(1) It is sometimes taken to be a version of ``Mach's principle" that ``the  spacetime metric is uniquely determined by the distribution of matter". And it is sometimes proposed that the principle can be captured  in the requirement that ``if one first specifies the energy-momentum distribution $T_{ab}$ on the spacetime manifold $M$, then there is exactly one (or at most one) Lorentzian metric $g_{ab}$ on $M$ that, together with $T_{ab}$, satisfies (\ref{EE regular})".  But there is a serious problem with the proposal. In general, one \emph{cannot} specify the energy-momentum  distribution in the absence of a spacetime metric.  Indeed, in typical cases the metric enters explicitly in the expression for $T_{ab}$. (Recall the expression (\ref{T-perfect fluid}) for a perfect fluid.) Thus, in looking for solutions to (\ref{EE regular}), one must, in general, solve simultaneously for the metric \emph{and} matter field distribution. 
 
  %
%
%

(2) Given any smooth metric $g_{ab}$ on $M$, there certainly exists a smooth symmetric field $T_{ab}$ on $M$ that, together with $g_{ab}$, is a solution to (\ref{EE regular}). It suffices to define $T_{ab}$ by the left side of the equation.  But the field $T_{ab}$ so introduced will not, in general, be the energy-momentum field associated with any known matter field.  It will not even satisfy the weak energy condition discussed in section \ref{matter fields}. With the latter constraint on $T_{ab}$ in place, Einstein's equation is an entirely non-trivial restriction on spacetime structure. 

Discussions of spacetime structure in classical relativity theory proceed on three levels according to the stringency of the constraints imposed on $T_{ab}$.  At the first level, one considers only ``exact solutions", i.e., solutions where  $T_{ab}$ is, in fact,  the aggregate energy-momentum field associated with one or more known matter fields. So, for  example, one  might undertake to find all perfect fluid solutions exhibiting particular symmetries.   At the second level, one considers  the larger class of what might  be called ``generic solutions", i.e., solutions where $T_{ab}$ satisfies one or more generic constraints (of which the weak and dominant energy conditions are examples). It is at this level, for example,  that the singularity theorems of Penrose and Hawking (Hawking and Ellis \shortcite{Hawking-Ellis}) are proved. Finally, at the third level, one drops all restrictions on $T_{ab}$, and Einstein's equation plays no role. Many results about global structure are proved at this level, e.g., the assertion that  there exist closed timelike curves in any relativistic spacetime $(M, g_{ab})$ where $M$ is compact.

(3) The role played by the cosmological constant in Einstein's equation remains a matter of controversy. Einstein initially added the term $(-\lambda g_{ab})$ in 1917 to allow for the possibility of a static cosmological model (which, at the time, was believed necessary to properly represent the actual universe).\footnote{He did so for other reasons as well (see Earman \shortcite{Earman/cosmologicalconstant}), but I will pass over them here.} But there were clear problems with doing so. In particular, one does not recover Poisson's equation (the field equation of Newtonian gravitation theory) as a limiting form of Einstein's equation unless $\lambda = 0$. (See point (4) below.) Einstein was quick to revert to the original form of the  equation after Hubble's redshift observations gave convincing evidence that the universe is, in fact, expanding. (That the theory suggested the possibility of cosmic expansion before those observations must count as one of its great successes.) Since then the constant has often been reintroduced to help resolve discrepancies between theoretical prediction and observation, and then abandoned when the (apparent) discrepancies were resolved.  The controversy continues.  Recent observations indicating an accelerating rate of cosmic expansion have led many cosmologists to believe that our universe is characterized by a positive value for $\lambda$. (See Earman \shortcite{Earman/cosmologicalconstant} for an overview.) 

Claims about the value of the cosmological constant are sometimes cast as claims about the ``energy-momentum content of the vacuum".  This involves bringing the term $(-\lambda g_{ab})$ from the left side of equation (\ref{EE regular}) to the right, and re-interpreting it as an energy-momentum field, i.e., taking Einstein's equation in the form
\begin{equation}\label{EE-vacuum}
R_{ab} \,  - \frac{1}{2} \, R \, g_{ab}   \,  = \,  8 \, \pi \, (T_{ab} +  T^{VAC}_{ab}), 
\end{equation}     
where  $ \displaystyle  T^{VAC}_{ab} = \frac{\lambda}{8 \pi} \, g_{ab}$. Here $T_{ab}$ is still understood to represent the aggregate energy-momentum of all normal matter fields. But  $T^{VAC}_{ab}$ is now understood to represent the residual energy-momentum associated with empty space.   Given any unit timelike vector $\xi^a$ at a point, $(T^{VAC}_{ab}\, \xi^a \, \xi^b)$ is $\displaystyle \frac{\lambda}{8 \pi}$.  So, on this re-interpretation, $\lambda$  comes out (up to the factor $8 \pi$) as the energy-density of the vacuum as determined by \emph{any} observer, at {any} point in spacetime. 

It should be noted that there is a certain ambiguity involved in referring to $\lambda$ as the cosmological \emph{constant} (and a corresponding ambiguity as to what counts as a solution to Einstein's equation).  We can take  $(M,  g_{ab}, T_{ab})$ to qualify if it satisfies the equation for some value (or other) of $\lambda$. Or, more stringently, one can take it to qualify if it satisfies the equation for some value of $\lambda$ that is fixed, once and for all, i.e., the same for all models  $(M,  g_{ab}, T_{ab})$.  In effect, we have here two versions of ``relativity theory". (See Earman \shortcite{Earman/cosmologicalconstant2} for discussion of what is at stake in choosing between the two.)

(4)  It is instructive to consider the  relation of Einstein's equation to Poisson's equation, the field equation of Newtonian gravitation theory:
\begin{equation} \label{Poisson without}
\nabla^2  \phi = 4 \, \pi \, \rho.
\end{equation}
Here $\phi$ is the Newtonian gravitational potential, and $\rho$ is the Newtonian mass density function.  In the ``geometrized"  formulation of the theory that we will consider in section \ref{Geometrized},  one trades in the potential $\phi$ in favor of a curved derivative operator, and Poisson's equation comes out as 
\begin{equation} \label{Poisson without geometric}
R_{ab} = 4 \, \pi  \rho \,  t_{ab},
\end{equation}
where $R_{ab}$ is the Ricci tensor field associated with the new curved derivative operator, and $ t_{ab}$ is the temporal metric.

The geometrized formulation of Newtonian gravitation was discovered after general relativity (in the 1920s).
But now, after the fact, we can put ourselves in the position of a hypothetical investigator who is considering possible candidates for a relativistic field equation, and knows about the geometrized formulation of Newtonian theory.  
What could be more natural than the attempt to adopt or adapt (\ref{Poisson without geometric})?  In the empty space case 
($\rho = 0$), this strategy suggests the equation $R_{ab} = \textbf{0}$, which is, of course, Einstein's equation (\ref{EE reversed}) for $T_{ab} = \textbf{0}$ and $\lambda = 0$.  This seems to me, by far, the best route to the latter equation.  Start with the Newtonian empty space equation ($R_{ab} = \textbf{0}$) and then simply leave it intact!

No such simple extrapolation is possible in the general case ($\rho \neq 0$). Indeed, I know of no heuristic argument for the full version of Einstein's equation (with or without cosmological constant) that is nearly so convincing.  But one can try something like the following.
The closest counterparts to (\ref{Poisson without geometric}) would seem to be ones of the form:  $R_{ab} = 4 \pi K_{ab}$, where $ K_{ab}$ is a symmetric tensorial function of $T_{ab}$ and $g_{ab}$.  The possibilities for $K_{ab}$ include   $T_{ab}$, \,  $g_{ab} \, T$,  \,  $T_{a}^{\ m}  T_{mb}$, \,  $g_{ab} \, (T^{mn} T_{mn})$, ...,  and linear combinations of these terms.    All but the first two involve terms that are second order or higher in $T_{ab}$. So, for example, in the special case of a dust field with energy density $\rho$ and four-velocity $\eta^a$,  they will  contain occurrences of $\rho^n$ with $n \geq 2$. (E.g., $g_{ab} \, (T^{mn} T_{mn})$ comes out as $\rho^2 g_{ab}$.)  But, presumably, only terms first order in $\rho$ should appear if the equation is to have a proper Newtonian limit.  This suggests that we look for a field equation of the form
\begin{equation} \label{candidate}
R_{ab} = 4 \, \pi  \left[k \, T_{ab} + l \,  g_{ab} T\right]
\end{equation}
or, equivalently,\footnote{Contraction on `$a$' and `$b$' in (\ref{candidate}) yields:  $R = 4 \, \pi  \,  (k + 4l) \, T$.  Solving for $T$, and substituting for $T$ in (\ref{candidate}) yields (\ref{equivalent candidate}).}  
\begin{equation} \label{equivalent candidate}
R_{ab}  - \frac{l}{(k+4l)}\,  R \,  g_{ab} =   4 \, \pi  \,  k  \, T_{ab},
\end{equation}
for some real  numbers  $k$ and $l$. Let $G_{ab}(k,l)$ be the field on the left side of the equation. It follows from the conservation condition that the field on the right side is divergence free, i.e.,  $\nabla_a \, ( 4 \, \pi  \,  k  \, T^{ab})  =  \mathbf{0}$.  So the conservation condition and (\ref{equivalent candidate}) can hold jointly only if
\begin{equation*}
\nabla_a \,  G^{ab}(k, l)  =  \mathbf{0}. 
\end{equation*} 
But by the ``Bianchi identity"(Wald \shortcite[pp.÷   39-40]{Wald}),  
\begin{equation} \label{Bianchi} 
\nabla_a \, (R^{ab}  - \frac{1}{2} \, R \, g^{ab})  = \, \mathbf{0}.  
\end{equation} 
The latter two conditions imply
\begin{equation*} 
\left[\frac{l}{(k + 4l)} - \frac{1}{2}\right]  \, \nabla_a (R g^{ab}) = \, \mathbf{0}.  
\end{equation*} 
Now $\nabla_a (R g^{ab}) = \, \mathbf{0}$ is an unreasonable constraint.\footnote{It implies that $R$ is constant and, hence, if (\ref{candidate}) holds, that $T$ is constant (since (\ref{candidate}) implies  $R = 4 \, \pi  \,  (k + 4 l) \, T$).  But this, in turn, is an unreasonable constraint on the energy-momentum distribution $T_{ab}$. E.g., in the case of a dust field with $T_{ab} = \rho \,  \eta^a \eta^b$,  $T = \rho$, and so the constraint implies that $\rho$ is constant. This is unreasonable since it rules out any possibility of cosmic expansion. (Recall the discussion toward the close of section \ref{matter fields}.)}  So the initial scalar term must be $0$. Thus, we are left with the conclusion that the conservation condition and (\ref{equivalent candidate}) can hold jointly only if $k + 2l = 0$, in which case (\ref{candidate}) reduces to 
\begin{equation} \label{candidate reduced}
R_{ab} = 4 \, \pi  \, k \left[T_{ab} -  \frac{1}{2} \,  g_{ab} \, T\right].
\end{equation}

It remains to argue that $k$ must be $2$ if (\ref{candidate reduced}) is to have a proper Newtonian limit. To do so, we consider, once again, the special case of a dust field with energy density $\rho$ and four-velocity $\eta^a$. Then, $T_{ab} = \rho \, \eta_a \, \eta_b$,  and  $T = \rho$. If we insert these values in (\ref {candidate reduced}) and contract with $\eta^a \, \eta^b$, we arrive at
\begin{equation} \label{Ricci contracted}
R_{ab} \, \eta^a \, \eta^b  =  2 \, \pi  \, k \, \rho.
\end{equation} 
Now the counterpart to a four-velocity field  in Newtonian theory is a vector field of unit temporal length, i.e., a field  $\eta^a$ where $t_{ab} \, \eta^a \, \eta^b = 1$.  If we contract  the geometrized version of Poisson's equation (\ref{Poisson without geometric}) with $\eta^a \, \eta^b$, we arrive at:   $R_{ab} \, \eta^a \, \eta^b  =  4 \, \pi  \, \rho$. Comparing this expression for  $R_{ab} \, \eta^a \, \eta^b$ with that in (\ref{Ricci contracted}), we are  led to the conclusion that $k = 2$, in which case (\ref{candidate reduced}) is just Einstein's equation (\ref{EE reversed}) with $\lambda = 0$.

Summarizing now, we have suggested that if one starts with the geometrized version of Poisson's equation (\ref{Poisson without geometric}) and looks for a relativistic counterpart, one is plausibly led to Einstein's equation with $\lambda = 0$. It is worth noting that if we had started instead with a variant of (\ref{Poisson without geometric}) incorporating a ``Newtonian cosmological constant"
\begin{equation} \label{Poisson without geometric + lambda}
R_{ab} + \lambda \, t_{ab} = 4 \, \pi  \rho \,  t_{ab},
\end{equation}
we would have been led instead to Einstein's equation (\ref{EE reversed}) without restriction on $\lambda$.  We can think of (\ref{Poisson without geometric + lambda}) as the geometrized version of 
\begin{equation} \label{Poisson without}
\nabla^2  \phi  + \lambda = 4 \, \pi \, \rho.
\end{equation}
\indent Let's now put aside the question of how one might try to motivate Einstein's equation. However one arrives at it, the equation -- let's now take it in the form (\ref{EE regular}) -- can be understood to assert a dynamical connection between a certain tensorial measure of spacetime curvature (on the left side) and the energy-momentum tensor field (on the right  side). It turns out that one can reformulate the connection in a way that makes reference only to scalar quantities, as determined relative to arbitrary observers.  The reformulation provides a certain insight into the geometric significance of the equation.\footnote{Another approach to its geometrical significance proceeds via the equation of geodesic deviation.  See, for example, Sachs and Wu \shortcite[p.÷  114]{Sachs-Wu/book}.}
  
Let $S$ be any smooth spacelike hypersurface in $M$.\footnote{We can take this to mean that $S$ is a smooth, imbedded, three-dimensional submanifold of $M$ with the property that any curve $\gamma:I \rightarrow M$ with image in $S$ is spacelike.}  The background metric  $g_{ab}$ induces a (three-dimensional) metric ${}^3g_{ab}$ on $S$.  In turn, this metric determines on $S$ a derivative operator, an associated Riemann curvature tensor field ${}^3R^a_{\ bcd}$, and a scalar curvature field ${}^3R =({}^3R^a_{\ bca})({}^3g^{bc})$.  Our reformulation  of Einstein's equation will direct attention to the values of ${}^3R$ at a point for a particular family of spacelike hypersurfaces passing through it.\footnote{In the case of a surface in three-dimensional Euclidean space, the associated Riemann scalar curvature ${}^2R$ is (up to a constant) just ordinary Gaussian surface curvature. We can think of ${}^3R$ in the present context as a higher dimensional analogue that  gives averaged values of Gaussian surface curvature. This can be made precise. See, for example, Laugwitz \shortcite[p.÷   127]{Laugwitz}.}  

\begin{figure}[h]
\begin{center}
\setlength{\unitlength}{1cm}
\begin{picture}(7.3,2.5)
 \put(0,0.0){\epsfig{figure=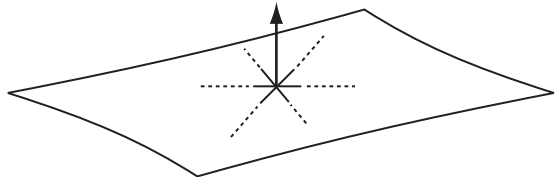}}
\end{picture} 
\begin{minipage}[t]{9.0cm}
\vspace{-1em}
\renewcommand{\baselinestretch}{1.0}
\caption{A ``geodesic hypersurface" through a point is constructed by projecting geodesics in all directions orthogonal to a given timelike vector there. } \label{geodesichypersurface}
\vspace{-.5em}
\end{minipage}
\end{center}
\end{figure}

Let $p$ be any point in $M$ and let $\xi^a$ be any future-directed unit timelike vector at $p$.  Consider the set of all geodesics through $p$ that are orthogonal to $\xi^a$ there. The (images of these) curves, at least when restricted to a sufficiently small open set containing $p$, sweep out a smooth spacelike hypersurface $S$.\footnote{More precisely, let  $S_p$ be the spacelike hyperplane in $M_p$ orthogonal to $\xi^a$. Then for any sufficiently small  open set $O$ in $M_p$ containing $p$, the image of $(S_p  \cap O)$  under the exponential map  $exp: O \rightarrow M$ is a smooth spacelike hypersurface. We can take it to be $S$.  (See, for example, Hawking and Ellis \shortcite[p.÷   33]{Hawking-Ellis}.) }  (See figure \ref{geodesichypersurface}.) 
\vspace{-.5em}
We will call  it a  \emph{geodesic hypersurface}.  (We cannot speak of \emph{the} geodesic hypersurface through $p$ orthogonal to $\xi^a$ because we have left open how far the generating geodesics are extended. But given any two, their restrictions to a suitably small open set containing $p$ coincide.) 

Geodesic hypersurfaces are of interest in their own right, the present context aside, because they are natural candidates for a notion of ``local simultaneity slice" (relative to a timelike vector at a point). 
What matters here, though,  is that, by the first Gauss-Codazzi equation (Wald \shortcite[p.÷  258]{Wald}), we have
\begin{equation} \label{first contraction Gauss-Codazzi}
{}^3R =  R \,  -2 \, R_{ab}  \xi^a  \xi^b 
\end{equation}
at $p$.\footnote{Let $\xi^a$ -- we use the same notation -- be the extension of the original vector at $p$ to a smooth future-directed unit timelike vector field on $S$ that is everywhere orthogonal to $S$. Then the first Gauss-Codazzi equation asserts that at \emph{all} points of $S$ 
\begin{equation*} 
{}^3R =  R \,  -2 \, R_{ab}  \xi^a  \xi^b \, + \pi_{ab}\, h^{ab}\,  + \,  \pi_{ab}\, \pi^{ab}, 
\end{equation*}
where $h_{ab}$ is the spatial projection field  $(g_{ab} - \xi_a \xi_b)$  on $S$, and $\pi_{ab}$ is the \emph{extrinsic curvature} field  $\frac{1}{2} \, \pounds_{\xi} h_{ab}$ on $S$.  But our construction guarantees that $\pi_{ab}$ vanish at $p$.} Here we have expressed the (three-dimensional) Riemann scalar curvature of $S$ at $p$ in terms of the (four-dimensional)  Riemann scalar curvature of $M$ at $p$ and the Ricci tensor there.  And so, if Einstein's equation  (\ref{EE regular}) holds, we have
\begin{equation} \label{Gauss-Codazzi+Einstein}
 {}^3R =  -16 \pi \,(T_{ab}\, \xi^a \xi^b) + 2 \lambda.  
\end{equation}
at $p$.

One can also easily work backwards to recover Einstein's equation at $p$ from the assumption that  (\ref{Gauss-Codazzi+Einstein}) holds for \emph{all} unit timelike vectors $\xi^a$ at $p$ (and all geodesic hypersurfaces through $p$ orthogonal to $\xi^a$). Thus, we have the following equivalence.

\begin{proposition} \label{E equivalent prop} 
Let $T_{ab}$ be a smooth symmetric field on $M$, and let $p$ be a point in $M$. Then Einstein's equation
$R_{ab} \,  - \frac{1}{2} \, R \, g_{ab}   + \lambda \, g_{ab} \,  = \,  8 \, \pi \, T_{ab}$
holds at $p$ iff for all future-directed unit timelike vectors $\xi^a$ at $p$, and all  geodesic hypersurfaces through $p$ orthogonal to $\xi^a$,  the scalar curvature ${}^3R$ of $S$ satisfies 
$ {}^3R =  [-16 \pi \,(T_{ab}\, \xi^a \xi^b) + 2 \lambda]$ at $p$.
\end{proposition}
The result is particularly instructive in the case where $\lambda = 0$. Then (\ref{Gauss-Codazzi+Einstein}) directly equates an intuitive scalar measure of spatial curvature (as determined relative to $\xi^a$) with energy density (as determined relative to $\xi^a$).  

\subsection{Congruences of Timelike Curves and ``Public Space"}

In this section, we consider congruences of timelike curves. We think of them as representing swarms of particles (or fluids).  First, we review the standard formalism for describing the local rotation and expansion of such congruences. Then, we consider different notions of ``space" and ``spatial geometry" as determined relative to them.

Once again, let $(M, \,   g_{ab})$  be our background relativistic spacetime (endowed with a temporal orientation).  Let $\xi^a$ be a smooth, future-directed, unit timelike vector field on $M$ (or some open subset thereof).  We understand it to be the four-velocity field  of our particle swarm.

Let $h_{ab}$ be the spatial projection field determined by $\xi^a$. Then the \emph{rotation} field $\omega_{ab}$ and the \emph{expansion} field $\theta_{ab}$ are defined as follows:
\begin{eqnarray}
\omega_{ab} & = & h_{[a}^{\ \ m} \, h_{b]}^{\ \ n} \, \nabla_m \, \xi_n \\
\theta_{ab} & = & h_{(a}^{\ \ m} \, h_{b)}^{\ \ n} \, \nabla_m \, \xi_n. 
\end{eqnarray}
\noindent  They are smooth fields, orthogonal to $\xi^a$ in both indices,  and satisfy
\begin{equation}\label{rot/exp decomposition 1}
\nabla_a \, \xi_b =   \omega_{ab} +  \theta_{ab}  +  \xi_a \, (\xi^n \nabla_n \,  \xi_b).  
\end{equation} 
\indent We can give the two fields $\omega_{ab}$ and $\theta_{ab}$ a geometric interpretation. Let $\eta^a$ be a vector field on the worldline of a particle $O$ that is ``carried along by the flow of $\xi^{a}$", i.e.,  $\pounds_{\xi} \, \eta^a = \textbf{0}$, and is orthogonal to $\xi^a$ at a point $p$. (Here $\pounds_{\xi} \, \eta^a$ is the Lie derivative of $\eta^a$ with respect to $\xi^a$.) We can think of $\eta^a$ at $p$ as a spatial ``connecting vector" that spans the distance between $O$ and a neighboring particle $N$ that is ``infinitesimally close".
%
%
\noindent The instantaneous velocity of $N$ relative to $O$ at $p$ is given by $\xi^n \, \nabla_n \, \eta^a$.  But $\xi^n \, \nabla_n \, \eta^a = \eta^n \, \nabla_n \, \xi^a $  (since $\pounds_{\xi} \, \eta^a = \textbf{0}$).  So, by (\ref{rot/exp decomposition 1}), and the orthogonality of $\xi^a$ with $\eta^a$ at $p$, we have 
\begin{equation}\label{rot/exp decomposition 2}
\xi^n \, \nabla_n \, \eta^a  =   (\omega_{n}^{\ \, a }  +  \theta_{n}^{\ \, a }) \, \eta^n.
\end{equation} 
\noindent at the point. Here we have simply decomposed the relative velocity vector into two components. The first,  $(\omega_{n}^{\ \, a } \, \eta^n)$, is orthogonal to $\eta^a$ (since $\omega_{ab}$ is anti-symmetric).  It  gives the instantaneous \emph{rotational  velocity} of  $N$ with respect to $O$ at $p$. 

In support of this interpretation, consider the instantaneous rate of change of the squared length $(-\eta^a \, \eta_a)$ of $\eta^a$ at $p$. It follows from (\ref{rot/exp decomposition 2}) that
\begin{equation}\label{rateofexpansion}
\xi^n \, \nabla_n \, (-\eta^a \, \eta_a)  = - \, 2 \, \theta_{na}\,  \eta^n \, \eta^a. 
\end{equation} 
\noindent Thus the computed rate of change depends solely on $\theta_{ab}$. Suppose $\theta_{ab} = \textbf{0}$. Then the instantaneous velocity of $N$ with respect to $O$ at $p$ has vanishing radial component.  If $\omega_{ab} \neq \textbf{0}$, $N$ still exhibits a non-zero velocity with respect to $O$. But it can only be a rotational velocity. The two conditions ($\theta_{ab} = \textbf{0}$ and \, $\omega_{ab} \neq \textbf{0}$) jointly characterize ``rigid rotation". 

The condition  $\omega_{ab} = \textbf{0}$, by itself,  characterizes \emph{irrotational}  flow. One gains considerable insight into the condition by considering a second, equivalent formulation.  Let us say that the field $\xi^a$ is \emph{hypersurface orthogonal} if there exist smooth, real valued maps $f$ and  $g$  (with the same domains of definition as $\xi^a$) such that, at all points, $\xi_a \, =  \, f \,  \nabla_a \,  g$.   Note that if the  condition is  satisfied, then the hypersurfaces of constant $g$ value are everywhere orthogonal to $\xi^a$.\footnote{For if $\eta^a$ is a vector tangent to one of these hypersurfaces, $\eta^n \nabla_n \, g   = 0$. So  $\eta^n  \xi_n = \eta^n  ( f \,  \nabla_n \, g) = 0$.}  Let us further say that  $\xi^a$ is \emph{locally hypersurface orthogonal} if the restriction of $\xi^a$ to every sufficiently small open set is hypersurface orthogonal.  
%
\begin{proposition}\label{Frobenius}
Let $\xi^a$ be a smooth, future-directed unit timelike vector field defined on $M$ (or some open subset of $M$). Then the following conditions  are equivalent.  
\nopagebreak[3]
\vspace{-.5em}
\begin{enumerate}
\vspace{-.5em}
\item [(1)] $\omega_{ab} = \textbf{0}$ everywhere. 
\vspace{-.5em}
\nopagebreak[3]
\item [(2)] $\xi^a$ is locally hypersurface orthogonal. 
\vspace{-.5em}
\end{enumerate}
\end{proposition}
The implication from (2) to (1) is immediate.\footnote{Assume that  $\xi_a \, =  \, f \,  \nabla_a \,  g$.  Then
\begin{eqnarray*}
\omega_{ab} & = &  h_{[a}^{\ \ m} \, h_{b]}^{\ \ n} \, \nabla_m \, \xi_n   =      h_{[a}^{\ \ m} \, h_{b]}^{\ \ n} \, \nabla_m \,  (f \, \nabla_n \, g)  \\
& = &  f \,   h_{[a}^{\ \ m} \, h_{b]}^{\ \ n} \, \nabla_m \,   \, \nabla_n \, g \  + \    h_{[a}^{\ \ m} \, h_{b]}^{\ \ n} \, (\nabla_m \, f) \, (\nabla_n \, g) \\
 & = &  f \,   h_a^{\ \ m} \, h_b^{\ \ n} \, \nabla_{[m} \,   \, \nabla_{n]} \, g \  + \    h_a^{\ \ m} \, h_b^{\ \ n} \, (\nabla_{[m} \, f) \, (\nabla_{n]} \, g). 
\end{eqnarray*}  
But $\nabla_{[m} \,   \, \nabla_{n]} \, g = \mathbf{0}$ since $\nabla_a$ is torsion-free,  and the second term in the final line vanishes as well since  $h_b^{\ \ n} \,  \nabla_n \, g = f^{-1} \,  h_b^{\ \ n} \, \xi_n = \mathbf{0}$. So $\omega_{ab} = \mathbf{0}$.}  But the converse is non-trivial. It is a special case of Frobenius's theorem (Wald \shortcite[p.÷   436]{Wald}).    The qualification `locally' can be dropped in (2) if the  domain of $\xi^a$ is, for example, simply connected.  

There is a nice picture that goes with the proposition. Think about an ordinary rope. In its natural twisted state, the rope cannot be sliced by an infinite family of slices in such a way that each  slice is orthogonal to all fibers. But if the rope is first untwisted, such a slicing is possible. Thus orthogonal sliceability is equivalent to fiber untwistedness. The proposition extends this intuitive equivalence to the four-dimensional ``spacetime ropes" (i.e., congruences of worldlines) encountered in relativity theory. It asserts that a congruence is irrotational (i.e., exhibits no twistedness) iff  it is, at least locally,  hypersurface orthogonal 

Suppose that our vector field $\xi^a$ \emph{is} irrotational and, to keep things simple,  suppose that its domain of definition is  simply connected. Then the hypersurfaces to which it is orthogonal are natural candidates for constituting ``space" at a given ``time"  relative to $\xi^a$ or, equivalently, relative to its associated set of integral curves.  This is a notion of \emph{public space} to be contrasted with \emph{private space},  which is determined relative to individual  timelike vectors or timelike curves.\footnote{The distinction between ``public space" and ``private space" is discussed in Rindler \shortcite{Rindler} and Page \shortcite{Page}. The terminology is due to E. A. Milne.}  Perhaps the best candidates for the latter are the ``geodesic hypersurfaces" we considered, in passing, in section \ref{Einstein's Equation}. (Given a point $p$ and a timelike vector $\xi^a$ there, we took a ``geodesic hypersurface through $p$ orthogonal to $\xi^a$" to be a spacelike hypersurface generated by geodesics through $p$ orthogonal  to $\xi^a$.)  

\begin{figure}[h]
\begin{center}
\setlength{\unitlength}{1cm}
\begin{picture}(4.55,2.55)
 \put(0,0.0){\epsfig{figure=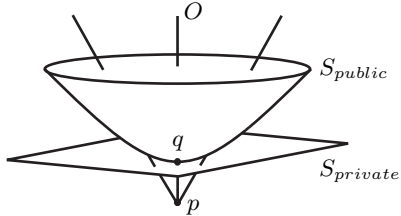}}
\put(2.35,2.5){\small $O$}
\put(4.15,.4){\small $S_{private}$}
\put(4.15,1.75){\small $S_{public}$}
\put(2.2,.8){\small $q$}
\put(2.4,-.05){\small $p$}
\end{picture} 
\begin{minipage}[b]{9.0cm}
\renewcommand{\baselinestretch}{1.0}
\caption{``Private space" $S_{private}$ at $p$ relative to $O$,  and ``public space" $S_{public}$  at $p$ relative to a congruence of timelike curves of which $O$ is a member.}\label{publicprivatespace}
\vspace{-.5em}
\end{minipage}
\end{center}
\end{figure}

The distinction between  public and private space is illustrated in Figure \ref{publicprivatespace}. 
There we consider a congruence of future-directed timelike half-geodesics in Minkowski spacetime starting at some particular point $p$.  One line $O$ in the congruence is picked out along with a  point $q$ on it. Private space relative to $O$ at $q$ is a spacelike hypersurface $S_{private}$ that is flat, i.e., the metric induced on $S_{private}$ has  a Riemann curvature tensor field ${}^3R^a_{\ bcd}$ that vanishes everywhere. In contrast,  public space at $q$ relative to the congruence is a spacelike hypersurface $S_{public}$ of constant negative curvature.  If $\xi^a$ is the future-directed unit timelike vector field everywhere tangent to the congruence, and $h_{ab} = (g_{ab} - \xi_a \, \xi_b)$  is its associated  spatial projection field,  then the curvature tensor field on $S_{public}$ associated with $h_{ab}$  has the form  ${}^3R_{abcd} \,  =  \,    -\frac{1}{K^2} (h_{ac} \, h_{bd} \, - \, h_{ad} \, h_{bc})$, where $K$ is the distance along $O$ from $p$ to $q$. (This is the characteristic form for a three-manifold of constant  curvature $-\frac{1}{K^2}$.)             

We have been considering ``public space" as determined relative to an irrotational  congruence of timelike curves. There is another sense in which one might want to use the term.  Consider, for example, ``geometry on the surface of a rigidly rotating disk" in Minkowski spacetime.  (There is good evidence that Einstein's  realization that this geometry is  non-Euclidean played an important role in his development of relativity theory (Stachel \shortcite{StachelRotatingDisk}).)  One needs to ask  in what sense the surface of a rotating disk \emph{has} a geometric structure. 

We can certainly model the rigidly rotating disk as a congruence of timelike curves in Minkowski spacetime. (Since the disk is two-dimensional, the congruence will be confined to a three-dimensional, timelike submanifold  $M'$ of $M$.)  But precisely because the disk is rotating, we cannot find hypersurfaces everywhere orthogonal to the curves and understand the geometry of the disk to be the geometry induced on them -- or, strictly speaking, induced on the two-dimensional manifolds determined by the intersection of the putative hypersurfaces with $M'$ -- by the background spacetime metric $g_{ab}$.

The alternative  is to think of ``space" as constituted by the ``manifold of trajectories", i.e., take the individual timelike curves in the congruence to play the role of spatial points, and consider the metric induced on \emph{this} manifold by the background spacetime metric.   The construction will not work for an arbitrary congruence of timelike curves.  It is essential that we are dealing here with a ``stationary" system.  (The metric induced on the manifold of trajectories (when the construction works) is fixed and frozen.) But it  \emph{does}  work for these systems, at least.  More precisely, anticipating the terminology of the following section, it works  if the four-velocity field of the congruence in question is proportional to a Killing field.  (The construction is presented in detail in Geroch \shortcite[Appendix]{GeneratingSolutions}.) 

Thus  we have two notions  of ``public space". One is available if the  four-velocity field of the congruence in question is irrotational; the other if it is proportional to a  Killing field. Furthermore,  if the four-velocity field is irrotational  \emph{and} proportional to Killing field, as is the case when we dealing with a ``static" system,  then the two notions of public space are essentially equivalent.

\subsection{Killing Fields and Conserved Quantities}

Let $\kappa^a$ be a smooth vector field on $M$. We say it is a \emph{Killing field} if $\pounds_{\kappa} \, g_{ab} = \textbf{0}$, i.e., if the Lie derivative with respect to $\kappa^a$ of the metric vanishes.\footnote{We drop the index on $\kappa$ here to avoid giving the impression that $\pounds_{\kappa} \, g_{ab} $ is a three index tensor field. Lie derivatives are always taken with respect to (contravariant) vector fields, so no ambiguity is introduced when the index is dropped.  A similar remark applies to our bracket notation below.}  This is equivalent to the  requirement that the ``flow maps" $\{\Gamma_s\}$ generated by $\kappa^a$ are all isometries. (See Wald \shortcite[p.÷   441]{Wald}.)  For this reason,  Killing fields are sometimes called ``infinitesimal generators of smooth one-parameter families of  isometries" or ``infinitesimal symmetries".   The defining condition can also be expressed as\footnote{This follows since $\pounds_{\kappa} \, g_{ab} = \kappa^n \nabla_n \,  g_{ab} + g_{nb} \nabla_a \, \kappa^n + g_{an} \nabla_b \, \kappa^n$,  and $\nabla_a$ is compatible with $g_{ab}$, i.e., $\nabla_n g_{ab} = \textbf{0}.$}
\begin{equation} \label{Killing's equation}
  \nabla_{(a} \, \kappa_{b)}  \, = \,  \textbf{0}. 
\end{equation}
\noindent  This is ``Killing's equation".

Given any two smooth vector fields $\xi^a$ and $\mu^a$ on $M$, the \emph{bracket} or \emph{commutator} field 
$[\xi ,  \mu]^a$  defined by $[\xi,   \mu]^a  =  \pounds_{\xi} \, \mu^a$  is also smooth.  The set of smooth vector fields on $M$ forms  a Lie algebra with respect to this operation, i.e., the bracket operation is linear in each slot; it is anti-symmetric ($[\xi,   \mu]^a  =  -[\mu,   \xi]^a$); and it satisfies the Jacobi identity
\begin{equation} \label{Jacoby identity}
 [[\xi ,  \mu],\, \nu]^a \, + \, [[\nu ,  \xi], \,  \mu]^a \, + \,  [[\mu ,  \nu], \, \xi]^a \, = \, \textbf{0}
\end{equation}
for all smooth vector fields $\xi^a$, $\mu^a$, and $\nu^a$ on $M$. It turns out that the bracket field of two Killing fields is also a Killing field.   So it follows, as well, that the  set of Killing fields on $M$ has a natural Lie algebra structure.    
 
The discussion of smooth symmetries in spacetime, and their associated conserved quantities,
is naturally cast in the language of Killing fields. For example, we can use that language to capture precisely the following intuitive notions.  A spacetime is \emph{stationary} if it has a Killing field that  is everywhere timelike. 
 It is \emph{axially symmetric} if it has a Killing field that is everywhere spacelike, and has integral curves that are closed.  (The ``axis" in this case is the set of points, possibly empty,  where the Killing field vanishes.) Finally, a spacetime is \emph{spherically symmetric} if it has three Killing fields $\overset{1}{\sigma}{}^a, \overset{2}{\sigma}{}^a, \overset{3}{\sigma}{}^a$ that (i) are everywhere spacelike, (ii) are linearly dependent at every point, i.e., 
$\overset{1}{\sigma}{}^{[a} \,  \overset{2}{\sigma}{}^b \, \overset{3}{\sigma}{}^{c]} = \textbf{0}$, and (iii) exhibit the same commutation relations as do the generators of the rotation group in three dimensions:  
\begin{equation} \label{Lie commutation relations}
[\overset{1}{\sigma}  \, , \,   \overset{2}{\sigma}]^ a \, = \, \overset{3}{\sigma}{}^a \, , \hspace{.3in}
[\overset{2}{\sigma}  \, , \,   \overset{3}{\sigma}]^ a \, = \, \overset{1}{\sigma}{}^a \, , \hspace{.3in}
[\overset{3}{\sigma}  \, , \,   \overset{1}{\sigma}]^ a \, = \, \overset{2}{\sigma}{}^a.
\end{equation}
 
Now we consider, very briefly, two types of conserved quantity. One is an attribute of massive point particles, the other of extended bodies.   Let $\kappa^a$ be an arbitrary Killing field, and let $\gamma: I \rightarrow M$ be a  timelike curve, with unit tangent field $\xi^a$, whose image is the worldline of a point particle with mass $m > 0$.  Consider the quantity  $J = (P^a  \kappa_a)$, where $P^a = m \,  \xi^a$ is the four-momentum of the particle.  It certainly need not be constant on $\gamma [I]$.  But it will be if  $\gamma$ is a geodesic. For in that case,  $\xi^n \nabla_n \, \xi^a = \textbf{0}$ and hence, by (\ref{Killing's equation}),  
\begin{equation} \label{1st Killing constancy equation}
\xi^n \nabla_n J =  m \,  (\kappa_a \, \xi^n \nabla_n \, \xi^a  +   \xi^n  \xi^a \, \nabla_n \, \kappa_a) \, = \,
m \,  \xi^n  \xi^a \, \nabla_{(n} \, \kappa_{a)} =  \,  \textbf{0}.   
\end{equation}
\noindent Thus, the value of $J$ (construed as an attribute of massive point particles) is constant for \emph{free} particles.       

We refer to $J$ as the conserved quantity associated with $\kappa^a$. If $\kappa^a$ is timelike, and if the flow maps  
determined by $\kappa^a$ have the character of translations\footnote{In Minkowski spacetime, one has an unambiguous classification of  Killing fields as generators of translations, spatial rotations, boosts (and linear combinations  of them).   No such classification is available in general. Killing fields are just Killing fields.  But sometimes a Killing field  in a curved spacetime resembles a Killing field in Minkowski spacetime in certain respects, and then the terminology may carry over naturally.  For example, in the case of asymptotically flat spacetimes, one can classify Killing fields by their asymptotic behavior.}, then $J$ is called the \emph{energy} of the particle (associated with  $\kappa^a)$.\footnote{If $\kappa^a$ is of unit length everywhere, this usage accords well with that in section \ref{Sp-Time Decomp}. For there ascriptions of energy to point particles were made relative to unit timelike vectors, and the value of the energy at any point was taken to be the inner product of that unit timelike vector with the particle's four-momentum vector. If $\kappa^a$ is, at least, of constant length, then one can always rescale it so as to achieve agreement of usage. But, in general, Killing fields, timelike or otherwise, are not of constant length, and so the current usage must be regarded as a generalization of that earlier usage.} If it is spacelike, and if the flow maps
have the character of translations, then $J$ is called the component of \emph{linear momentum} of the particle (associated with  $\kappa^a$). Finally,  if $\kappa^a$ is spacelike, and if the flow maps have the character of rotations, then it is called the component of \emph{angular momentum} of the particle (associated with  $\kappa^a$). 

\begin{figure}[h]
\begin{center}
\setlength{\unitlength}{1cm}
\begin{picture}(4.59,5.3)
 \put(0,0.0){\epsfig{figure=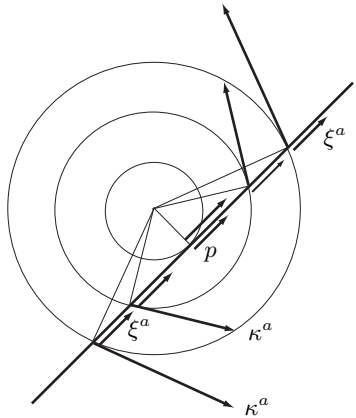}}
\put(2.62,2.0){\small  $p$}
\put(3.2,0.90){\small $\kappa^a$}
\put(3.15,-.1){\small $\kappa^a$}
\put(4.2,3.5){\small $\xi^a$}
\put(1.60,0.95){\small $\xi^a$}
\end{picture} 
\begin{minipage}[t]{9.0cm}
\vspace{-1em}
\renewcommand{\baselinestretch}{1.0}
\caption{$\kappa^a$ is a rotational Killing field. (It is everywhere orthogonal to a circle radius, and proportional to it in length.) $\xi^a$ is a tangent vector field of constant length on the line. The inner-product between them is constant. (Equivalently, the length of the projection of $\kappa^a$ onto the line is  constant.)} \label{EuclideanKilling}
\vspace{-.5em}
\end{minipage}
\end{center}
\end{figure}

It is useful to keep in mind a certain picture that helps one to ``see" why the angular momentum of free particles (to take that  example)  is conserved. It involves an analogue of angular momentum in Euclidean plane geometry. Figure  \ref{EuclideanKilling} 
shows a rotational Killing field $\kappa^a$ in the Euclidean plane, the image of a geodesic (i.e., a line $L$), and the tangent field $\xi^a$ to the geodesic.   
Consider the quantity $J = \xi^a  \kappa_a$, i.e., the inner product of $\xi^a$ with $\kappa^a$,  along $L$.  Exactly the same proof as before (in equation (\ref{1st Killing constancy equation})) shows that $J$ is constant along $L$.\footnote{The mass $m$ played no special role.} But here we can better visualize the  assertion.  

Let us temporarily drop indices and write $\kappa \cdot \xi$ as one would in ordinary Euclidean vector calculus (rather than $ \xi^a  \kappa_a$). Let $p$ be the point on $L$ that is closest to the center point where $\kappa$ vanishes. At that point,  $\kappa$ is parallel to $\xi$. As one moves away from $p$ along $L$, in either direction,  the length $|\kappa|$ of
$\kappa$ grows, but the angle $\angle(\kappa, \xi)$ between the vectors increases as well.   It is at least plausible from the picture (and easy to check directly with an argument involving similar triangles)  that the length of the projection of $\kappa$ onto the line is constant. Equivalently,  the inner product $\kappa \cdot \xi =  cos (\angle(\kappa, \xi))   \, \|\kappa\| \, \|\xi\|$ is constant. 

That is how to think about the conservation of angular momentum for  free particles in relativity theory.  It does not matter that in the latter context we are dealing with a Lorentzian  metric and allowing for curvature. The claim is still that a certain inner product of  vector fields remains constant along a geodesic, and we can still think of that constancy as arising from a compensatory balance of two factors. 

Let us now turn to the second type of conserved quantity, the one that is an attribute of extended bodies. Let $\kappa^a$ be an arbitrary Killing field, and let $T_{ab}$ be the energy-momentum field associated with some matter field. Assume it satisfies the conservation condition.  Then $(T^{ab} \, \kappa_b)$ is divergence free:
\begin{equation} \label{2nd Killing constancy equation}
\nabla_a (T^{ab}  \kappa_b)  \, = \,  \kappa_b \, \nabla_a T^{ab} \, + \, T^{ab} \nabla_a \kappa_b \, = \, T^{ab} \nabla_{(a} \kappa_{b)} \, = \, \textbf{0}.  
\end{equation}
(The second equality follows from the conservation condition for $T^{ab}$ (in section \ref{matter fields})  and the symmetry of $T^{ab}$; the third from the fact that $\kappa^a$ is a Killing field.)  It is natural, then, to apply Stokes' theorem to the vector field $(T^{ab} \kappa_b)$. 

Consider a bounded system with aggregate energy-momentum field $T_{ab}$ in an otherwise empty universe.  Then  there exists a (possibly huge) timelike world tube such that $T_{ab}$ vanishes outside the tube (and vanishes on its boundary).  
\begin{figure}[h]
\begin{center}
\setlength{\unitlength}{1cm}
\begin{picture}(7.3,4.4)
 \put(0,0.0){\epsfig{figure=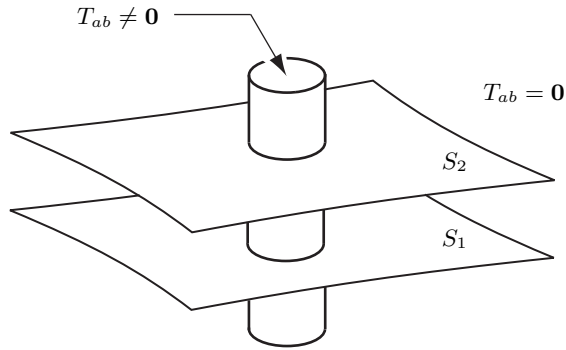}}
\put(5.75,1.35){\small  $S_1$}
\put(5.75,2.38){\small  $S_2$}
\put(6.3,3.3){\small $T_{ab} = \mathbf{0}$}
\put(0.9,4.30){\small $T_{ab}  \neq \mathbf{0}$}
\end{picture} 
\begin{minipage}[t]{9.0cm}
\vspace{-1em}
\renewcommand{\baselinestretch}{1.0}
\caption{The integrated energy (relative to a background timelike Killing field) over the intersection of the world tube with a spacelike hypersurface is independent of the choice of hypersurface.} \label{globalconservation}
\vspace{-.5em}
\end{minipage}
\end{center}
\end{figure}
Let $S_1$ and $S_2$ be (non-intersecting) spacelike hypersurfaces that cut the tube as in figure \ref{globalconservation}, and let $N$ be the segment of the tube falling between them (with boundaries included). By Stokes' theorem,
\begin{eqnarray*} \label{Killing integral equation}
\lefteqn{\int_{S_2}  (T^{ab} \kappa_b) \,  dS_a  - \int_{S_1}  (T^{ab} \kappa_b) \, dS_a }  \hspace{1in}\\
  & =& \int_{S_2 \cap \,  \partial N}  (T^{ab} \kappa_b) \,  dS_a  - \int_{S_1 \cap \,   \partial N}  (T^{ab} \kappa_b) \, dS_a \\ 
& =& \int_{\partial N}  (T^{ab} \kappa_b) \,  dS_a   
=   \int_{N} \,  \nabla_a  (T^{ab} \kappa_b) \, dV = 0.\end{eqnarray*}
Thus, the integral \  $\int_{S}  (T^{ab} \kappa_b) \,  dS_a$ \ is independent of the choice of spacelike hypersurface $S$ intersecting the world tube, and is, in this sense, a conserved quantity (construed as an attribute of the system confined to the  tube).   An ``early" intersection yields the same value as a ``late" one.  Again, the character of the background Killing field $\kappa^a$ determines our description of the conserved quantity in question.  If $\kappa^a$ is timelike, we take \  $\int_{S}  (T^{ab} \kappa_b) \,  dS_a$ \ to be the aggregate energy of the system  (associated with  $\kappa^a$). And so forth. 

   For further discussion of symmetry and conservation principles in general relativity, see Brading and Castellani (this volume, chapter 13).

\section{Special Topics}

\subsection{Relative Simultaneity in Minkowski Spacetime}\label{simultaneity}

We noted in section \ref{Sp-Time Decomp}, when discussing the decomposition of vectors at a point into their ``temporal" and ``spatial" components relative to a four-velocity vector there, that we were taking for granted the standard identification  of relative simultaneity with orthogonality.  Here we return to consider the justification of that identification.

Rather than continue to cast the discussion as one concerning the decomposition of the tangent space at a particular point, it is  convenient to construe it instead as one about the structure of Minkowski spacetime, the regime of so-called ``special relativity".    Doing so will bring it closer to the framework in which traditional discussions of the status  of the relative simultaneity relation have been conducted.
 
 \emph{Minkowski spacetime} is a relativistic spacetime $(M, g_{ab})$ characterized by three conditions:  (i) $M$ is the manifold $\mathbb{R}^4 $;  (ii) $(M, g_{ab})$ is flat, i.e.,  $g_{ab}$ has vanishing  Riemann curvature everywhere; and (iii)  $(M, g_{ab})$ is geodesically complete, i.e., every geodesic (with respect to $g_{ab}$) can be extended to arbitrarily large parameter values in both directions.  

By virtue of these conditions, Minkowski spacetime can be canonically identified with its tangent space at any point, and so it inherits  
the structure of a ``metric affine space" in the following sense.  Pick any point $o$ in $M$,  and let $V$ be the tangent space $M_o$ at $o$.  Then there is a map $(p, q) \mapsto  \overrightarrow{pq}$ \, from $M \times M$ to $V$ with the following two properties.
\vspace{-.5em}
\begin{enumerate}
\item [(1)]  For all $p$, $q$ and $r$ in $M$, $\overrightarrow{pq} + \overrightarrow{qr} = \overrightarrow{pr}$.
\item[(2)]  For all $p$ in $M$, the induced map  $q \mapsto  \overrightarrow{pq}$ from $M$ to $V$ is a bijection.\footnote{If  $exp$ is the exponential map from $M_o$ to $M$, we can take $\overrightarrow{pq}$ to be the vector 
\[
(exp^{-1}(q) - exp^{-1}(p))
 \] 
 in $M_o$. All other standard properties  of affine spaces follow from these two. E.g., it follows that \  $\overrightarrow{pq} = \mathbf{0} \  \Longleftrightarrow \  p = q$, for all  $p$ and $q$ in $M$. (Here $\mathbf{0}$ is the zero vector in $V$.)}
\end{enumerate}

The triple consisting of the point set $M$, the vector space $V$, and the map \,  $(p, q) \mapsto  \overrightarrow{pq}$ \, forms an \emph{affine space}.  If we add to this triple the inner product on $V$ defined by $g_{ab}$ it becomes a (Lorentzian)  \emph{metric affine space}. (For convenience we will temporarily drop the index notation and write  $\langle v, w\rangle$ instead of  $g_{ab} v^a w^b$ for $v$ and $w$ in $V$.) We take all this structure for granted in what follows, i.e., we work with Minkowski  spacetime and construe it as a metric affine space in the sense described. This will simplify the presentation considerably. 
 
 
We also use an obvious notation for orthogonality. Given four points $p,  q,  r, s$  in $M$, we write \,  $\overrightarrow{pq} \perp \overrightarrow{rs}$ \,  if  \, $\langle \overrightarrow{pq},  \overrightarrow{rs} \rangle = 0$.  And given a line\footnote{In the present context we can characterize a \emph{line} in more than one way. We can take it to be the  image of a maximally extended geodesic that is non-trivial, i.e., not a point. Equivalently, we can take it to be a set of points of the form   $\{r:  \overrightarrow{pr} =  \epsilon \, \overrightarrow{pq}$ for some $\epsilon$ in  $\mathbb{R}\}$ where $p$ and $q$ are any two (distinct) points in $M$.} $O$ in $M$,  we write \,  $\overrightarrow{pq} \perp O  $ \, if \, $\overrightarrow{pq} \perp \overrightarrow{rs}$  for all points $r$ and $s$ on $O$.    


Now consider a timelike line $O$ in $M$. What pairs of points $(p, q)$ in  $M$ should qualify as being ``simultaneous relative to $O$"?  That is the question we are considering.    The standard answer is that they should do so precisely if $\overrightarrow{pq} \perp O$.  

\begin{figure}[h]
\begin{center}
\setlength{\unitlength}{1cm}
\begin{picture}(2.1,4.4)
 \put(0,0.0){\epsfig{figure=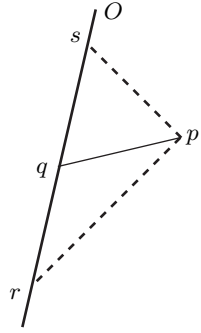}}
\put(2.2,2.5){\small  $p$}
\put(.2,2.05){\small  $q$}
\put(-.15,.38){\small  $r$}
\put(.65,3.8){\small  $s$}
\put(1.1,4.1){\small  $O$}
\end{picture} 

\begin{minipage}[b]{9.0cm}
\renewcommand{\baselinestretch}{1.0}
\caption{The  $\epsilon = \frac{1}{2}$ characterization of relative simultaneity: $p$ and $q$ are simultaneous relative to $O$ iff $q$ is midway between $r$ and $s$.}\label{epsilon}  
\vspace{-.5em}
\end{minipage}
\end{center}
\end{figure}

 In traditional discussions of relative simultaneity, the standard answer is often cast in terms of ``epsilon" values. The connection is easy to see.  Let $p$ be any point that is not on our timelike line $O$.   Then there exist unique points $r$ and $s$ on $O$ (distinct from one another) such that $\overrightarrow{rp}$ and  $\overrightarrow{ps}$ are future-directed null vectors.  (See figure \ref{epsilon}.) 
Now let $q$ be any point on $O$. (We think of it as a candidate for being judged simultaneous with $p$ relative to $O$.) Then
\, $\overrightarrow{rq}  =  \epsilon \, \overrightarrow{rs}$ \,   for some $\epsilon\in \mathbb{R}$. A simple computation\footnote{First note that, since $\overrightarrow{ps}$ and $\overrightarrow{pr}$  are null,
\begin{equation*}
0 \, = \,    \langle\overrightarrow{ps},  \overrightarrow{ps} \rangle \, = \,  \langle\overrightarrow{pr} + \overrightarrow{rs}, \,  \overrightarrow{pr}+ \overrightarrow{rs} \rangle  \, = \,  2 \langle\overrightarrow{pr},  \overrightarrow{rs} \rangle + \langle\overrightarrow{rs},  \overrightarrow{rs} \rangle.     
\end{equation*}  
It follows that
\begin{equation*}
\langle\overrightarrow{pq},  \overrightarrow{rs} \rangle \, = \,  \langle\overrightarrow{pr} +  \overrightarrow{rq}, \,  \overrightarrow{rs} \rangle  \, = \,\langle\overrightarrow{pr} +  \epsilon \overrightarrow{rs}, \,  \overrightarrow{rs} \rangle  \, = \,   \langle\overrightarrow{pr},  \overrightarrow{rs} \rangle +  \epsilon \langle\overrightarrow{rs},  \overrightarrow{rs} \rangle \, = \, (\epsilon - \frac{1}{2})  \langle\overrightarrow{rs},  \overrightarrow{rs}\rangle,   
\end{equation*}  
which implies (\ref{epsilon-orthgonality}).}
shows that 
\begin{equation} \label{epsilon-orthgonality}
\epsilon = \frac{1}{2} \, \Longleftrightarrow \, \overrightarrow{pq} \perp \overrightarrow{rs} . 
\end{equation}
So the standard (orthogonality) relation of relative simultaneity in special relativity may equally well  be described as the ``$\epsilon = \frac{1}{2}$" relation of relative simultaneity.

Yet another equivalent formulation involves  the ``one-way speed of light".  Suppose a light ray travels from $r$ to $p$ with speed $c_{+}$ relative to $O$, and from $p$ to $s$ with speed $c_{-}$ relative to $O$. We saw in section \ref{Sp-Time Decomp} that \emph{if} one adopts the standard criterion of relative simultaneity, then it follows that $c_{+} = c_{-}$.  The converse is true as well.  For if  $c_{+} = c_{-}$, then, as determined relative to $O$, it should take as much time for light to travel from $r$ to $p$ as from $p$ to $s$. And in that case, a point $q$ on $O$ should be judged simultaneous with $p$ relative to $O$ precisely if it is midway between $r$ and $s$. So we are led, once again,  to the ``$\epsilon = \frac{1}{2}$" relation of relative simultaneity.   

\emph{Now is adoption of the standard relation a matter of convention,  or is it in some significant sense forced on us?}

 
There is, of course, a large literature devoted to this question.\footnote{Classic statements of the conventionalist position can be found in Reichenbach  \shortcite{Reichenbach} and Gr\"unbaum \shortcite{Grunbaum-classic}.  Gr\"unbaum has recently responded to criticism of his views in \shortcite{Grunbaum-new}. An overview of the debate with many references can be found in Janis \shortcite{Janis}. }  It is not my purpose to review it here,  but I do want to draw attention to certain remarks of Howard Stein \shortcite[pp.÷   153-4]{Stein/contraMaxwell} that seem to me particularly insightful. 
He makes the point that determinations of  conventionality require a context.  
 \begin{quote} 
There are really two distinct aspects to the issue of the ``conventionality" of Einstein's concept  of relative simultaneity. One may assume the position of Einstein himself at the outset of his investigation -- that is, of one confronted by a problem, trying to find a theory that will deal with it satisfactorily; or one may assume the position of (for instance) Minkowski -- that is, of one confronted with a theory already developed, trying to find its most adequate and instructive formulation.
\end{quote}  


The problem Einstein confronted was (in part) that of trying to account for our apparent inability to detect any motion of the earth with respect to the ``aether". A crucial element of his solution was the proposal that we think about simultaneity a certain way (i.e., in terms of  the ``$\epsilon = \frac{1}{2}$ criterion"),   and resolutely follow through on the consequences of doing so. Stein emphasizes  just how different that proposal looks when we consider it, not from Einstein's initial position, but rather from the vantage point of the finished theory, i.e., relativity theory conceived as an account of invariant spacetime structure. 
\begin{quote}
\begin{sloppypar} 
[For] Einstein, the question (much discussed since Reichenbach)  whether the evidence really shows that that the speed of light \emph{must} be regarded as the same in all directions and for all observers is not altogether appropriate. A person devising a theory does not have the responsibility, at the outset, of showing that the theory being developed is the only possible one given the evidence. [But] once Einstein's theory had been developed, and had proved successful in dealing with all relevant phenomena, the case was quite transformed; for we know that \emph{within} this theory, there is only one ``reasonable" concept of simultaneity (and in terms of that concept, the velocity of light is indeed as Einstein supposed); therefore an alternative will only present itself if someone succeeds in constructing, not simply a different empirical criterion of simultaneity, but an essentially different (and yet viable) theory of electrodynamics of systems in motion. No serious alternative theory is in fact known.  (emphasis in original)
\end{sloppypar} 
\end{quote}  

My goal in the remainder of this  section is to formulate three elementary uniqueness results, closely related to one another, that capture the sense in which ``there is only one `reasonable' concept of (relative) simultaneity" within the framework of Minkowski spacetime.  

It will help to first consider an analogy.  In some formulations of Euclidean plane geometry,  the relation of congruence between angles is taken as primitive along with that of congruence between line segments (and other relations suitable for formulating axioms about affine structure).  But suppose we have a formulation in which it is not, and we undertake to \emph{define} a notion of angle-congruence in terms of the other primitives.  The standard angle-congruence relation can certainly be defined this  way, and there is a clear sense in which it is the only reasonable candidate.  Consider any two angles in the Euclidean plane.  (Let's agree that an ``angle" consists of two rays, i.e.,  half-lines, with a common initial point.)  Whatever else is the case, presumably, it is only reasonable to count them as congruent, i.e., equal in ``size",   if there is an isometry of the Euclidean plane  that maps one angle onto the other.\footnote{In this context, a one-to-one map of  the Euclidean plane onto itself is an ``isometry"  if it preserves the relation of congruence between line segments.}  So though we have here a notion of angle-congruence that is introduced ``by definition", there is no interesting sense in which it is conventional in character.

A situation very much like this arises if we think about  ``one-way light speeds" in terms of Minkowskian spacetime geometry. Indeed, the claim that the speed of light \emph{in vacuo} is the same in all directions and for all inertial observers is naturally represented as a claim about angle congruence (for a special type of angle) in Minkowski spacetime.  

\begin{figure}[h]
\begin{center}
\setlength{\unitlength}{1cm}
\begin{picture}(5.25,2.7)
 \put(0,0.0){\epsfig{figure=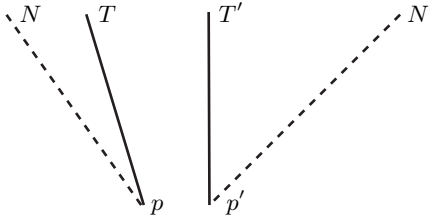}}
\put(1.95,-.05){\small  $p$}
\put(2.95,-.05){\small  $p'$}
\put(.20,2.45){\small  $N$}
\put(1.25,2.45){\small  $T$}
\put(5.35,2.45){\small  $N'$}
\put(2.85,2.45){\small  $T'$}
\end{picture} 
\begin{minipage}[b]{9.0cm}
\renewcommand{\baselinestretch}{1.0}
\caption{Congruent ``light speed angles" in Minkowski spacetime.}\label{lightspeedangle}  
\vspace{-.5em}
\end{minipage}
\end{center}
\end{figure}

Let us take a ``light-speed angle" to be a triple of the form $(p, T, N)$, where $p$ is a point in $M$, $T$ is a future-pointing timelike ray with initial point $p$,  and $N$ is a future-pointing null ray with  initial point $p$.   (See figure \ref{lightspeedangle}.)
Then we can represent systematic attributions of one-way light speed  as maps of the form:  $(p, T, N) \mapsto v(p, T, N)$.  (We understand  $v(p, T, N)$ to be the speed that an observer with (half) worldline $T$ at $p$ assigns to the light signal with (half) worldline $N$.)  So, for example, the principle that the speed of light is the same in all directions and for all inertial observers comes out as the condition that  $v(p, T, N)  =  v(p', T', N')$ for all light-speed angles 
$(p, T, N)$ and $(p', T', N')$.

Now it is natural to regard  $v(p, T, N)$  as a measure of the ``size" of the angle $(p, T, N)$.  If we do so, then, just as in the Euclidean case, we can look to the background metric to decide when two  angles have the same size.  That is, we can take them to be congruent iff there is an isometry of Minkowski spacetime that maps one to the other.   But on this criterion, all light-speed angles are congruent (proposition \ref{lightspeedangles}).  So we are led back to the principle that the (one-way) speed of  light is the same in all directions and for all inertial observers and, hence, back to the standard relative simultaneity relation. 
\begin{proposition} \label{lightspeedangles} 
Let $(p, T, N)$ and $(p', T', N')$ be any two light speed angles in Minkowski spacetime. Then there is an isometry $\phi$ of Minkowski spacetime such that $\phi(p) = p'$,  $\phi[T] = T'$, and   $\phi[N] = N'$.\footnote{The required isometry can be realized in the form $\phi = \phi_3 \circ \phi_2 \circ \phi_1$ where  (i)   $\phi_1$ is a translation that  takes $p$ to $p'$, (ii)  $\phi_2$ is a  boost (based at $p'$) that maps $\phi_1[T]$ to $T'$, and (iii) $\phi_3$ is a rotation about $T'$ that maps $(\phi_2 \circ \phi_1)[N]$ to $N'$.}
\end{proposition}


Once again, let $O$ be a timelike line in $M$, and let $Sim_O$ be the standard relation of simultaneity relative to $O$. (So $(p,q)\in Sim_O$ iff  $\overrightarrow{pq} \perp O$, for all $p$ and $q$ in $M$.)  Further, let $S$  be an arbitrary two-place relation on $M$ that we regard as a candidate for the relation of ``simultaneity relative to O". Our second uniqueness result asserts that if $S$ satisfies three conditions, including an invariance condition,   then $S = Sim_O$.\footnote{Many propositions of this form can be found in the literature.  (See Budden \shortcite{Budden} for a review.)  Ours is intended only as an example.  There are lots of possibilities here depending on exactly how one formulates the conditions that $S$ must satisfy. The proofs are all very much the same.}  

The first two conditions are straightforward.
\vspace{-.5em}
\begin{enumerate}
\item [S1] $S$ is an equivalence relation (i.e., $S$ is reflexive, symmetric, and transitive). 
\item[S2]  For all points $p \in M$, there is a unique point $q \in O$ such that  $(p, q) \in S$.
 \end{enumerate}
	
\noindent If $S$ satisfies (S1),  it has an associated family of equivalence classes.  We can think of them as ``simultaneity slices" (as determined relative to $O$). Then  (S2) asserts that every simultaneity slice intersects $O$ in exactly one point. 

The third condition  is intended to capture  the requirement that $S$ is determined by the background geometric structure of Minkowski spacetime and by $O$ itself.  But there is one subtle point here. It makes a difference whether  temporal orientation counts as part of that background geometric structure or not.  Let's assume for the moment that it does not. 

Let us say, quite generally,  that  $S$ is \emph{implicitly definable} in a structure of the form $(M, \,  ...)$ if it is invariant under all symmetries of  $(M, \,  ... )$, i.e., for all such symmetries $\phi:M \rightarrow M$, and all points $p$ and $q$ in $M$,
\begin{equation} \label{implicit definability}
(p, q) \in  S  \ \Longleftrightarrow  \  (\phi(p), \phi(q))\in  S.
\end{equation}
(Here, a \emph{symmetry} of $(M, \,  ... )$ is understood to be a bijection $\phi:M \rightarrow M$ that preserves all the structure in ``...".) 
This is certainly a very weak sense of definability.\footnote{We follow Budden \shortcite{Budden} in using  ``implicit definability in $(M, ... )$"  as a convenient device for organizing various closely related  uniqueness results. One moves from one to another simply by shifting the choice  of ``..." . }

We are presently interested in the structure  $(M, \,   \langle \  \rangle, \,  O)$.\footnote{Strictly speaking,  since we are here thinking of Minkowski spacetime as a metric affine space,  we should  include elements of structure (to the right of $M$) that turn the point set $M$ into an affine space, i.e., add a four-dimensional vector space $V$ and a map from $M \times M$ to $V$ satisfying conditions (1) and (2) listed at beginning of this section. But it is cumbersome to do so every time. Let them be understood in what follows.}  Its symmetries are bijections  $\phi:M \rightarrow M$ such that,  for all $p, q, r, s$ in $M$, 
\begin{equation}
\langle  \overrightarrow{\phi(p) \, \phi(q)}, \ \overrightarrow{\phi(r) \, \phi(s)}   \rangle = \langle \overrightarrow{pq}, \ \overrightarrow{rs}   \rangle
\end{equation} 
and
\begin{equation}
p \in O  \ \Longleftrightarrow \ \phi(p) \in O. 
\end{equation} 
 They are generated by maps of the following three types:  (i) translations (``up and down")  in the direction of $O$, (ii) spatial rotations that leave fixed every point in $O$, and  (iii) temporal reflections with respect to spacelike hyperplanes orthogonal to $O$. Our second uniqueness result can be formulated as follows.\footnote{It is a close variant of one presented in Hogarth \shortcite{Hogarth}.} 
\begin{proposition} \label{1st-S-uniqueness} 
Let $S$ be a two-place relation on Minkowski spacetime that satisfies conditions (S1), (S2), and is implicitly definable in $(M, \,   \langle \  \rangle, \,  O)$. Then $S$ = $Sim_O$.
\end{proposition}
As it turns out, the full strength of the stated  invariance condition is not needed here.  It suffices to require that $S$ be invariant under maps of type (iii).\footnote{The key step in the proof is the following.  Let $p$  be a point in $M$. By (S2), there is a unique point $q$ on $O$ such that $\langle p, q \rangle \in S$. Let $\phi: M \rightarrow M$  be a reflection with respect to the hyperplane orthogonal to $O$ that passes through $p$.  Then $\phi(p) = p$, $\phi(q) \in O$, and $S$ is invariant under $\phi$.  Hence\ $ \langle p, \phi(q) \rangle  \, = \,  \langle \phi(p), \phi(q) \rangle \in S$.  Since $\phi(q) \in O$, it follows by the uniqueness condition in (S2) that $\phi(q) = q$. But the only points left fixed by $\phi$ are those on the  hyperplane orthogonal to $O$ that passes through $p$. So $p$ and $q$ are both on that hyperplane, and  $\overrightarrow{pq}$ is  orthogonal to $O$, i.e.,  $\langle p, q \rangle \in Sim_O.$}

Suppose now that we \emph{do} want to consider  temporal orientation as part of the background structure that may play a role in the determination of $S$. Then  the requirement of implicit definability in $(M, \,   \langle \  \rangle, \,  O)$  should be replaced by that of implicit definability in $(M, \,  \mathcal{T}, \,  \langle \  \rangle, \,  O)$,  where $\mathcal{T}$ is the background temporal orientation.  Maps of type (i) and (ii) qualify as symmetries of this enriched structure (too), but temporal reflections of type (iii) do not. Implicit definablity in $(M, \, \mathcal{T},    \langle \  \rangle, \,  O)$ is a weaker condition than  implicit definability in $(M, \,   \langle \  \rangle, \,  O)$, and  proposition \ref{1st-S-uniqueness}  fails if the latter condition is replaced by the former. 

But we can still get a uniqueness result if we change the set-up  slightly:  namely, we think of $O$ as merely one member of a congruence of parallel timelike lines $\mathcal{F}_O$, the \emph{frame} of $O$, and think of simultaneity as determined relative to the latter. Implicit definability in $(M, \, \mathcal{T},  \,   \langle \  \rangle, \,  \mathcal{F}_O)$ is a stronger condition than implicit definability in $(M, \, \mathcal{T}, \,   \langle \  \rangle, \,  O)$ because there are symmetries of the former  -- (iv) translations taking $O$ to other lines in $\mathcal{F}_O$, and  (v) spatial rotations that leave fixed the points of some line in $ \mathcal{F}_O$ other than $O$  --  that are not symmetries of the latter.  Our variant formulation of the uniqueness result is the following.\footnote{It is closely related to propositions in Spirtes \shortcite{Spirtes}, Stein \shortcite{Stein/contraMaxwell}, and Budden \shortcite{Budden}.}
\begin{proposition} \label{2nd-S-uniqueness} 
Let $S$ be a two-place relation on Minkowski spacetime that satisfies conditions (S1), (S2), and is implicitly definable in $(M, \, \mathcal{T},  \,   \langle \  \rangle, \,  \mathcal{F}_O)$. Then $S$ = $Sim_O$.
\end{proposition}

The move from proposition \ref{1st-S-uniqueness} to proposition  \ref{2nd-S-uniqueness} involves a trade-off.  We drop the requirement that $S$ be invariant under maps of type (iii), but add the requirement that it be invariant under those of type (iv) and (v).\footnote{It is a good exercise to check that one does not need the full strength of the stated invariance condition to derive the conclusion. It suffices to require that $S$ be invariant under maps of type (i) and (v), or, alternatively, invariant under maps of type (ii) and (iv).}

Once again, many variations of these results can be found in the literature.   
For example, if one subscribes to a ``causal theory of time (or spacetime)", one will want to consider what candidate simultaneity relations are determined by the causal structure of Minkowski spacetime (in adddition to the line $O$).   Let $\kappa$ be the symmetric two-place relation of ``causal connectibility" in $M$, i.e., the relation that holds of two points $p$ and $q$ if \,  $\overrightarrow{pq}$ \,  is a causal vector.  Clearly, every symmetry of $(M, \langle  \ \rangle)$ is a symmetry of  $(M, \kappa)$. So the requirement of implicit definability in $(M, \,   \kappa, \,  O)$ is at least as strong as that of implicit definability in  $(M, \,   \langle  \ \rangle, \,  O)$.  It follows that we can substitute the former for the latter in proposition \ref{1st-S-uniqueness}. Similarly, we can substitute the requirement of implicit definability in $(M, \, \mathcal{T},  \,   \kappa,  \,  \mathcal{F}_O)$ for that of implicit definability in $(M, \, \mathcal{T},  \,   \langle  \ \rangle,  \,  \mathcal{F}_O)$ in proposition \ref{2nd-S-uniqueness}.

\subsection{Geometrized Newtonian Gravitation Theory}\label{Geometrized}

The ``geometrized" formulation  of Newtonian gravitation theory was first introduced by Cartan \shortcite{Cartan-1923,Cartan-1924},  and Friedrichs \shortcite{Friedrichs}, and later developed by Trautman  \shortcite{Trautman/lectures},   K\"unzle \shortcite{Kunzle,Kunzle2}, 
Ehlers \shortcite{Ehlerslimits}, and others. 

It is significant for several reasons. (1)  It shows that several features of relativity theory once thought to be uniquely characteristic of it  do not distinguish it from (a suitably reformulated version of) Newtonian gravitation theory. The latter too can be cast as a ``generally covariant" theory in which  (a) gravity emerges as a manifestation of spacetime curvature, and (b) spacetime structure is ``dynamical",  i.e., participates in the unfolding of physics rather than being a fixed backdrop against which it unfolds.  

(2) It helps one to  see where Einstein's equation ``comes from", at least in the empty-space case. (Recall the discussion in section \ref{Einstein's Equation}.) It also allows one to make precise, in coordinate-free, geometric language,   the  standard claim that Newtonian gravitation theory (or, at least, a certain generalized version of it) is the ``classical limit" of general relativity.  (See  K\"unzle \shortcite{Kunzle2} and Ehlers \shortcite{Ehlerslimits}.)  

(3) It clarifies the gauge status of the Newtonian gravitational potential.  In the geometrized formulation of Newtonian theory,  one works with a single curved derivative operator $\stackrel{g \hspace{.4em}}{\nabla_a}$. It can be decomposed (in a sense) into two pieces -- a flat derivative operator $\nabla_a$ and a gravitational potential $\phi$ -- to recover the standard formulation of the theory.\footnote{As understood here,  the  ``standard" formulation is not that found in undergraduate textbooks, but rather a ``generally covariant"  theory of four-dimensional spacetime structure in which gravity is not geometrized.} But in the absence of special boundary conditions, the decomposition will not be unique.  Physically, there is no unique way to divide into ``inertial" and ``gravitational"components the forces experienced by particles. Neither has any direct physical significance. Only their ``sum" does. It is an attractive feature of  the geometrized formulation that it trades in two gauge quantities for this sum.

(4) The clarification  described in (3) also leads to  a solution, or dissolution, of an old conceptual problem about Newtonian gravitation theory,  namely the apparent breakdown of the theory when applied (in cosmology) to a hypothetically infinite, homogeneous mass distribution. (See Malament \shortcite{Malament-Newtcosmology} and Norton \shortcite{Norton-Newtcosmology,Norton-cosmologicalwoes}.)

In what follows, we give a brief overview of the geometrized formulation of Newtonian gravitation theory, and say a bit more about points (1) and (3).  We start by characterizing a new class of geometrical models for the spacetime structure of our universe (or subregions thereof) that is broad enough to include the models considered in both the standard and geometrized versions of Newtonian theory.  We take a \emph{classical spacetime} to be a structure  $(M, t_{ab}, h^{ab}, \nabla_a)$ where (i) $M$ is a smooth, connected, four-dimensional differentiable manifold; (ii) $t_{ab}$ is a smooth, symmetric, covariant tensor field on M of signature (1, 0, 0, 0);\footnote{The stated signature condition is equivalent to the requirement that, at every point $p$ in $M$, the tangent space $M_p$ has a basis \ $\overset{_1}{\xi} \ \! \!^a ,...,\overset{_4}{\xi} \ \!\!^a$  \ such that, for all $i$ and $j$ in $\{1,2,3,4\}$,
\[
t_{ab} \, \overset{_i}{\xi} \, \!^a  \overset{_i}{\xi} \,\!^b   = \left\{
\begin{array}{ll}
1 & \mbox{if  \ $i=1$}\\
0 & \mbox{if \  $i = 2, \, 3, \,  4$}
\end{array}
\right.
\]
and
$t_{ab} \, \overset{_i}{\xi} \, \!^a  \overset{_j}{\xi} \,\!^b   = 0$ if  \ $i\ne j$. Similarly, the signature condition on $h^{ab}$ stated in (iii) requires that, at every point, the co-tangent space have a basis $\overset{_1}{\sigma} \ \! \!_a ,...,
\overset{_4}{\sigma} \ \!\!_a$  such that, for all $i$ and $j$ in $\{1,2,3,4\}$,
\[
h^{ab} \, \overset{_i}{\sigma} \, \!_a  \overset{_i}{\sigma} \,\!_b   = \left\{
\begin{array}{ll}
0 & \mbox{if  \ $i=1$}\\
1 & \mbox{if \  $i = 2, \, 3, \,  4$}
\end{array}
\right.
\]
and
$h^{ab} \, \overset{_i}{\sigma} \, \!_a  \overset{_j}{\sigma} \,\!_b   = 0$ if  \ $i\ne j$.} (iii) $h^{ab}$ is a smooth, symmetric, contravariant tensor field on $M$ of signature $(0,ÊÊ1, 1, 1)$;   (iv) $\nabla_a$  is a smooth derivative operator on $M$; and (v) the following two conditions are met:
\begin{eqnarray}
h^{ab} \, t_{bc} &=& \mathbf{0}  \label{orthogonality condition}  \\
\nabla_a \,  t_{bc} &=& \textbf{0}   \hspace{.7em}     = \hspace{.7em}  \nabla_a \, h^{bc} .    \label{compatibility condition}
\end{eqnarray}
We refer to them, respectively, as the ``orthogonality" and ``compatibility" conditions.

 $M$ is interpreted as the manifold of point events (as before);   $t_{ab}$  \, and  \,  $h^{ab}$ are understood to be temporal and spatial metrics on $M$, respectively; and  $\nabla_a$  is understood to be an affine structure on $M$. Collectively, these objects represent the spacetime structure presupposed by classical, Galilean relativistic dynamics. We review, briefly,  how they do so.

In what follows, let $(M, t_{ab}, h^{ab}, \nabla_a)$ be a fixed classical spacetime.

Consider, first, $t_{ab}$.  Given any vector $\xi^a$ at a point,  it assigns a ``temporal length" $(t_{ab} \,  \xi^a  \, \xi^b)^{\frac{1}{2}}  \geq  0$.  The vector $\xi^a$ is classified as \emph{timelike} or \emph{ spacelike} depending on whether its temporal length is positive or zero.  It follows from the signature of $t_{ab}$ that the subspace of spacelike vectors at any point is three-dimensional. It also follows from the signature that at every point there exists a covariant vector $t_a$, unique up to sign, such that  $t_{ab} = t_a  t_b$.  We say that the structure  $(M, t_{ab}, h^{ab}, \nabla_a)$  is \emph{temporally orientable} if there is a continuous (globally defined) vector field $t_a$ such that this decomposition holds at every point.  Each such field $t_a$ (which, in fact, must be smooth because $t_{ab}$ is) is a \emph{temporal orientation}.  A timelike vector $\xi^a$ qualifies as \emph{future-directed}  relative to $t_a$ if  $ t_a \, \xi^a  > 0$; otherwise it is \emph{past-directed}. 
Let us assume in what follows that  $(M, t_{ab}, h^{ab}, \nabla_a)$ is temporally orientable and that a temporal orientation $t_a$ has been selected.

From the compatibility condition, it follows that $t_a$ is closed, i.e., $\nabla_{[a} \, t_{b]} = \mathbf{0}$.  So, at least locally, it must be exact, i.e., of form $t_a = \nabla_a \, t$ for some smooth function $t$. We call any such function a  \emph{time function}.  If $M$ has a suitable global structure, e.g., if it is simply connected, then a globally  defined time function  $t:M\rightarrow \mathbb{R}$ \,  must exist.  In this case, spacetime can be decomposed into a one-parameter family of global ($t$ = constant) ``time slices". One can speak of ``space" at a given ``time". A different choice of time function would result in a different zero-point for the time scale, but would induce the same time slices and the same elapsed intervals between them. 


We say that a smooth curve is  \emph{timelike} (respectively \emph{spacelike}) if  its tangent field is timelike (respectively spacelike) at every point.   In what follows, unless indication is  given to the contrary, it should further be understood that  a ``timelike curve" is future-directed and parametrized by its $t_{ab}$ - length.  In this case, its tangent field $\xi^a$ satisfies the normalization condition $t_a \xi^a = 1$. Also, in this case,  if a particle happens to have the image of the curve as its worldline,  then, at any point, $\xi^a$ is called the particle's \emph{four-velocity}, and $\xi^n \nabla_n \, \xi^a$ its \emph{four-acceleration}, there.\footnote{Here we take for granted an interpretive principle that corresponds to $C1$:  (i) a curve is timelike iff its image could be the worldline of a point particle. Other principles we can formulate at this stage correspond to $P1$ and $P2$:  (ii) a timelike curve can be reparametrized so as to be a geodesic (with respect to $\nabla_a $) iff its image could be the worldline of a free particle;  (iii) clocks record the passage of elapsed $t_{ab}$ - length along their worldlines.  (Here, in contrast the relativistic setting, we have only massive particles to consider and, until we geometrize Newtonian gravity, do not count a particle as ``free" if it is subject to ``gravitational force".)} If the particle has mass $m$,  then its  four-acceleration field satisfies the equation of motion
\begin{equation} \label{Newt second law again}
F^a = m \ \xi^n \nabla_n \, \xi^a, 
\end{equation}
where $F^a$ is a spacelike vector field (on the image of its worldline) that represents the net force acting on the particle. This is, once again, our version of Newton's second law of motion. Recall  (\ref{Second Law}).  Note that the equation makes geometric sense because four-acceleration vectors are necessarily spacelike.\footnote{By the compatibility condition,     $t_a \, \xi^n \nabla_n \, \xi^a =  \xi^n \nabla_n \, (t_a \,\xi^a) = \xi^n \nabla_n \, (1) = 0$.}

Now consider $h^{ab}$. It serves as a spatial metric, but just how it does so is a bit tricky.  In Galilean relativistic mechanics,  we have no notion of spatial length for timelike vectors, e.g., four-velocity vectors,  since having one is tantamount to a notion of absolute rest.  (We can take a particle to be at rest if its four-velocity has spatial length $0$ everywhere.) But we \emph{do} have a notion of spatial length for spacelike vectors, e.g., four-acceleration vectors.  (We can, for example, use measuring rods to determine distances between simultaneous events.) $h^{ab}$ serves to give us one without the other.

We cannot take the  spatial length of a vector $\sigma^a$ to be $(h_{ab} \, \sigma^a  \sigma^b )^{\frac{1}{2}}$ because the latter is not well-defined. (Since $h^{ab}$ has degenerate signature, it is not invertible, i.e., there does not \emph{exist} a field $h_{ab}$ satisfying  $h^{ab} h_{bc} = \delta^{a}_{\hspace{.4em}c}$.) But if $\sigma^a$ is spacelike, we can use $h^{ab}$ to  assign a spatial length to it indirectly.  It turns out that:  (i) a vector $\sigma^a$ is spacelike iff it can be expressed in the form $\sigma^a = h^{ab} \, \lambda_b$, and (ii) if it can be so expressed, the quantity
$(h^{ab} \, \lambda_a \, \lambda_b)$  is independent of the choice of $\lambda_a$. 
Furthermore, the signature of $h^{ab}$ guarantees that  $(h^{ab} \, \lambda_a \, \lambda_b) \geq 0$.  So if $\sigma^a$ is spacelike, we can take its spatial length to be $(h^{ab} \, \lambda_a \, \lambda_b )^{\frac{1}{2}}$, for any choice of corresponding  $\lambda_a$.     



One final preliminary remark about classical spacetimes is needed. It is crucial for our purposes, as will be clear,  that the compatibility condition (\ref{compatibility condition}) does not determine a unique derivative operator. (It is a fundamental result that the compatibility condition $\nabla_a \, g_{bc} = \mathbf{0}$ determines a unique derivative operator if $g_{ab}$ is a semi-Riemannian metric, i.e., a smooth, symmetric field that is invertible (i.e., non-degenerate). But neither  $t_{ab}$ nor $h^{ab}$ is invertible.)

Because $h^{ab}$ is not invertible, we cannot raise \emph{and} lower indices with it. But we can, at least, raise indices with it, and it is sometimes convenient to do so.  So, for example, if $R^a_{\ bcd}$ is the Riemann curvature tensor field associated with $\nabla_a$, we can understand  $R^{ab}_{\ \ cd}$ to be an abbreviation for  $h^{bn} \, R^{a}_{\ ncd}$.   
   
\begin{sloppypar}
Let us now, finally, consider Newtonian gravitation theory.  In the standard (non-geometrized) version, one works with a flat derivative operator $\nabla_a$ and a gravitational potential $\phi$, the latter understood to be a smooth, real-valued function on $M$. The gravitational force on a point particle with mass $m$ is given by  $- \, m \, h^{ab} \, \nabla_b \, \phi$.   (Notice that this is a spacelike vector by the orthogonality condition.) Using our convention for raising indices, we can also express the vector as:   $- \, m \, \nabla^a \, \phi$.   It follows that  if the particle is subject to no forces except gravity, and if it has four-velocity $\xi^a$,  it satisfies the equation of motion    
\begin{equation}  \label{mass free equation of motion}
- \nabla^a \, \phi  \, = \, \xi^n \, \nabla_n \, \xi^a.
\end{equation}	
(Here we have just used  $- \, m \, \nabla^a \, \phi $ for the left side of (\ref{Newt second law again}).) It is also assumed that $\phi$  satisfies Poisson's equation:
\begin{equation}
\nabla^a \, \nabla_a  \, \phi  \,  = \,   4  \, \pi  \, \rho,	
\end{equation}	
where $\rho$ is the Newtonian mass-density function (another smooth real-valued function on $M$).  (The expression on the left side is an abbreviation for:  $h^{ab} \, \nabla_b \, \nabla_a  \, \phi $.)
\end{sloppypar}

In the geometrized formulation of the theory, gravitation is no longer conceived as a fundamental ``force" in the world, but rather as a manifestation of spacetime curvature (just as in relativity theory). Rather than thinking of point particles as being deflected from their natural straight (i.e., geodesic) trajectories, one thinks of them as traversing geodesics in curved spacetime. So we have a geometry problem. Starting with the structure $(M, t_{ab}, h^{ab}, \nabla_a)$, can we find a new derivative operator $\stackrel{g \hspace{.4em}}{\nabla_a}$, also compatible with the metrics $t_{ab}$ and $h^{ab}$, such that a timelike curve satisfies the equation of motion (\ref{mass free equation of motion}) with respect to the original derivative operator $\nabla_a$  iff it is a geodesic with respect to $\stackrel{g \hspace{.4em}}{\nabla_a}$? The following proposition (essentially due to Trautman \shortcite{Trautman/lectures}) asserts that there is exactly one such $\stackrel{g \hspace{.4em}}{\nabla_a}$. It also records several facts about the Riemann curvature tensor field $\stackrel{g}{R}{}\hspace{-.2em}^a{}_{bcd}$ associated with $\stackrel{g \hspace{.4em}}{\nabla_a}$.  

In formulating the proposition,  we make use of the following basic fact about derivative operators. Given any two such operators $\stackrel{1 \hspace{.5em}}{\nabla_a}$ and $\stackrel{2 \hspace{.5em}}{\nabla_a}$  on $M$, there is a unique smooth tensor field $C^a_{\ bc}$, symmetric in its covariant indices, such that, for all smooth fields  $\alpha^{a \ldots b}_{\hspace{1.5em} c \ldots d}$  on $M$,
\begin{eqnarray} 
\stackrel{2 \hspace{.5em}}{\nabla_n } \alpha^{a \ldots b}_{\hspace{1.5em} c \ldots d}&=&\stackrel{1 \hspace{.5em}}{\nabla_n } \alpha^{a \ldots b}_{\hspace{1.5em} c \ldots d} \, + \,  C^r_{\ nc} \,  \alpha^{a \ldots b}_{\hspace{1.5em} r \ldots d} \, 
+ \,  ...  \, + \, C^r_{\ nd} \,  \alpha^{a \ldots b}_{\hspace{1.5em} c \ldots r}  \nonumber  \\
&  & \hspace{2em} {} - C^a_{\ nr} \,  \alpha^{r \ldots b}_{\hspace{1.5em} c \ldots d} \, 
- \,  ...  \,  - C^b_{\ nr} \,  \alpha^{a \ldots r}_{\hspace{1.5em} c \ldots d}.      \label{C expansion}
\end{eqnarray}
In this case, we say that ``the action of $\stackrel{2 \hspace{.5em}}{\nabla_a}$ relative to that of $\stackrel{1 \hspace{.5em}}{\nabla_a}$ is given by $C^a_{\ bc}$".\footnote{Clearly, if the action of $\stackrel{2 \hspace{.5em}}{\nabla_a}$ relative to that of $\stackrel{1 \hspace{.5em}}{\nabla_a}$ is given by $C^a_{\ bc}$, then, conversely, the action of $\stackrel{1 \hspace{.5em}}{\nabla_a}$ relative to that of $\stackrel{2 \hspace{.5em}}{\nabla_a}$ is given by $-C^a_{\ bc}$.
In the sum on the right side of (\ref{C expansion}),  there is one term involving $C^a_{\ bc}$  for each index in $\alpha^{a \ldots b}_{\hspace{1.5em} c \ldots d}$. In each case, the index in question is contracted with $C^a_{\ bc}$, and the term carries a coefficient of $+1$ or $-1$ depending on whether the index in question is in covariant (down) or contravariant (up) position.  (The components of $C^a_{\ bc}$ in a particular coordinate system are obtained by subtracting the Christoffel symbols associated with $\stackrel{1 \hspace{.5em}}{\nabla_a}$ (in that coordinate system) from those associated with $\stackrel{2 \hspace{.5em}}{\nabla_a}$.)} Conversely, given any one derivative operator $\stackrel{1 \hspace{.5em}}{\nabla_a}$  on $M$,  and any smooth, symmetric field $C^a_{\ bc}$Êon $M$, (\ref{C expansion}) defines a new derivative operator $\stackrel{2 \hspace{.5em}}{\nabla_a}$ on $M$. (See Wald \shortcite[p.÷   33]{Wald}.)
\begin{proposition}  [Geometrization Theorem] \label{Geometrization Theorem} Let $(M, t_{ab}, h^{ab}, \nabla_a)$ be a classical spacetime with $\nabla_a$ flat ($R^a{}_{bcd} = \mathbf{0}$).  Further,  let  $\phi$ and $\rho$  be smooth real valued functions on $M$ satisfying Poisson's equation: $\nabla^a \, \nabla_a  \, \phi  =  4  \, \pi  \, \rho.$   Finally, let  $\stackrel{g \hspace{.4em}}{\nabla_a}$ be the derivative operator on $M$ whose action relative to that of $\nabla_a$ is given by $C^a_{\ bc}  =  -t_{bc} \, \nabla^a \phi$.  Then all the following hold. \\
\indent (G1)  $(M, t_{ab}, h^{ab}, \stackrel{g \hspace{.4em}}{\nabla_a})$  is a classical spacetime. \\
\indent (G2)  $\stackrel{g \hspace{.4em}}{\nabla_a}$  is the unique derivative operator on $M$ such that, for all timelike \\
\indent  \hspace{2em} curves on $M$ with four-velocity fields $\xi^a$, 
\vspace{-.5em}
\begin{equation}  \label{geometrization condition}
\xi^n  \stackrel{g \hspace{.4em}}{\nabla_n}  \xi^a \, = \,  \mathbf{0}  \  \Longleftrightarrow  \ 
- \nabla^a \phi  \, = \, \xi^n \, \nabla_n \, \xi^a. 
\vspace{-.5em}  
\end{equation}
\indent (G3) The curvature field $\stackrel{g}{R}{}\hspace{-.2em}^a{}_{bcd}$ associated with $\stackrel{g \hspace{.4em}}{\nabla_a}$ satisfies: 
\vspace{-.6 em}
\begin{alignat}{2} 
&\stackrel{g \hspace{.3em}}{R_{bc}}  &= \ \     &4 \, \pi \,   \rho \,  t_{bc}   \label{geometric Poisson}\\
&\stackrel{g}{R}{}\hspace{-.2em}^{ab}_{\ \ cd}   &= \ \    &\mathbf{0}  \label{first integrability condition} \\
&\stackrel{g}{R}{}\hspace{-.2em}{}^{[a}{}_{(b}{}^{c]}{}_{d)} \   &= \ \      &\mathbf{0}.  \label{second integrability condition} 
\end{alignat} 
\end{proposition}

(\ref{geometric Poisson}) is the geometrized version of Poisson's equation. The proof proceeds by more-or-less straight forward computation using (\ref{C expansion}).\footnote{Here is a sketch. By (\ref{C expansion}),
\begin{equation*}
\stackrel{g \hspace{.4em}}{\nabla_a}  t_{bc} = 
\nabla_a \, t_{bc} + C^r_{\ ab}\, t_{rc} + C^r_{\ ac}\, t_{br} = \nabla_a \, t_{bc}  +  ( -t_{ab} \, \nabla^r \phi)\, t_{rc} + ( -t_{ac} \, \nabla^r \phi)\, t_{br}. 
\end{equation*}
The first term in the far right sum vanishes by the compatibility condition (\ref{compatibility condition}); the second and third do so by the orthogonality condition (\ref{orthogonality condition}) since, for example,  $(\nabla^r \phi)\, t_{br} = (h^{rm} \,  t_{br}) \,\nabla_m \phi$.  So $\stackrel{g \hspace{.4em}}{\nabla_a}$ is compatible with $t_{bc}$. Much the same argument shows that it is also compatible with $h^{ab}$. This give us (G1). 

For (G2), let  $\stackrel{g \hspace{.4em}}{\nabla_a}$ (temporarily) be an arbitrary derivative operator on $M$ whose action relative to that $\nabla_a$ is given by some field $C^a_{\ bc}$.  Let $p$ be an arbitrary point in $M$, and let $\xi^a$ be the four-velocity field of an arbitrary timelike curve through $p$. Then, by (\ref{C expansion}), 
\begin{equation*}
\xi^n \stackrel{g \hspace{.4em}}{\nabla_n} \xi^a \,  = \,  \xi^n \nabla_n  \, \xi^a - C^a_{\ rn} \, \xi^r  \xi^n.  
\end{equation*}
It follows that $\stackrel{g \hspace{.4em}}{\nabla_n}$ will satisfy (G2) iff    $C^a_{\ rn} \xi^r  \xi^n  = -\nabla^a \phi$ or, equivalently, 
\begin{equation*}
[C^a_{\ rn} \,  + (\nabla^a \phi) \, t_{rn}] \,  \xi^r   \xi^n  = \mathbf{0}, 
\end{equation*}
for all future-directed unit timelike vectors $\xi^a$ at all points $p$.  But the space of  future-directed unit timelike vectors at any $p$ spans the tangent space $M_p$ there, and the field in brackets is symmetric in its covariant indices.  So, $\stackrel{g \hspace{.4em}}{\nabla_n}$ will satisfy (G2) iff  $C^a_{\ rn} \,  = \,   -(\nabla^a \phi) \, t_{rn}$  everywhere.

Finally, for (G3) we use the fact that  $\stackrel{g}{R}{}\hspace{-.2em}^a{}_{bcd}$ can be expressed as a sum of terms involving $R^a_{\ bcd}$ and $C^a_{\ bc}$ (see Wald \shortcite[p.÷   184]{Wald}), and then substitute for $C^a_{\ bc}$: 
\begin{eqnarray*}
\stackrel{g}{R}{}\hspace{-.2em}^a{}_{bcd} &=&  R^a_{\ bcd}  \, + \, 2 \,  \nabla_{[c} \, C^a_{\  d]b} \, + \, 2 \,   C^n_{\  b[c}  C^a_{\  d]n}  \nonumber \\
{}&=&  R^a_{\ bcd}  \, - \, 2 \,  t_{b[d}  \nabla_{c]} \,  \nabla^a \phi  \, = \, - \, 2 \, t_{b[d}  \nabla_{c]} \,  \nabla^a \phi . 
\end{eqnarray*}
(Here $C^n_{\  b[c}  C^a_{\  d]n}$ turns out to be $\mathbf{0}$ by the orthogonality condition, and  $\nabla_{[c} \, C^a_{\  d]b}$ turns out to be $- t_{b[d}  \nabla_{c]} \,  \nabla^a \phi$ by the compatibility condition. For the final equality we use our assumption that  $R^a_{\ bcd} = \mathbf{0}$.) (\ref{first integrability condition})  and (\ref{second integrability condition}) now follow from  the orthogonality condition and (for (\ref{second integrability condition})) from the fact that $\nabla^{[c} \nabla^{a]} \phi  = \mathbf{0}$ for \emph{any} smooth function $\phi$. Contraction on `$a$' and `$d$' yields  
\begin{equation*}
\stackrel{g}{R}{}\hspace{-.2em}_{bc} =    t_{bc}  (\nabla_{a}  \nabla^a \phi).    
\end{equation*}
So (\ref{geometric Poisson}) follows from our assumption that $\nabla^a \, \nabla_a  \, \phi  =  4  \, \pi  \, \rho$ (and the fact that $\nabla_a \,  \nabla^a \phi = \nabla^a \,  \nabla_a \phi$).}

We can also work in the  opposite direction.  In geometrized Newtonian gravitation theory,  one  \emph{starts} with a curved derivative operator  $\stackrel{g \hspace{.4em}}{\nabla_a}$ satisfying (\ref{geometric Poisson}), (\ref{first integrability condition}), (\ref{second integrability condition}), and with the principle that point particles subject to no forces (except ``gravity") traverse geodesics with respect to  $\stackrel{g \hspace{.4em}}{\nabla_a}$.   (\ref{first integrability condition}) and (\ref{second integrability condition}) function  as integrability conditions that ensure the possibility of working backwards to recover the standard formulation in terms of a gravitational potential and flat derivative operator.\footnote{I am deliberately passing over some subtleties here. Geometrized Newtonian gravitation theory comes in several variant formulations.   (See Bain \shortcite{Bain} for a careful review of the differences.)   The one presented here is essentially that of Trautman  \shortcite{Trautman/lectures}.  In other weaker formulations (such as that in K\"unzle \shortcite{Kunzle}),  condition (\ref{first integrability condition}) is dropped, and it is \emph{not} possible to fully work back to the standard formulation (in terms of a gravitational potential and flat derivative operator) unless special global conditions on spacetime structure are satisfied. 
} We have the following recovery, or de-geometrization,  theorem (also essentially due to Trautman \shortcite{Trautman/lectures}).


\begin{proposition} [Recovery Theorem] \label{Recovery Theorem}  Let $(M, t_{ab}, h^{ab}, \stackrel{g \hspace{.4em}}{\nabla_a})$ be a classical spacetime that, together with a smooth, real-valued function $\rho$ on $M$, satisfies conditions  (\ref{geometric Poisson}), (\ref{first integrability condition}), (\ref{second integrability condition}).  Then, at least locally (and globally if $M$ is, for example, simply connected), there exists a smooth, real-valued function $\phi$ on $M$ and a flat derivative operator $\nabla_a$ such that all the following hold. \\
\indent (R1)  $(M, t_{ab}, h^{ab}, \nabla_a)$  is a classical spacetime. \\
\indent (R2)    For all timelike curves on $M$ with four-velocity fields $\xi^a$,  the geometri-  \\
\indent  \indent  \indent  zation condition (\ref{geometrization condition})  is satisfied.  \\
\indent (R3) $\nabla_a$ satisfies Poisson's equation:  $\nabla^a \, \nabla_a  \, \phi  =  4  \, \pi  \, \rho.$
 \end{proposition}
The theorem is an existential assertion of this form: given $\stackrel{g \hspace{.4em}}{\nabla_a}$ satisfying certain conditions, there exists (at least locally) a smooth function  $\phi$ on $M$ and a flat derivative operator $\nabla_a$ such that  $\stackrel{g \hspace{.4em}}{\nabla_a}$ arises as the ``geometrization" of the pair  $(\nabla_a, \phi)$. But, as claimed at the beginning of this section,
we do \emph{not} have uniqueness unless special boundary conditions are imposed on $\phi$.  

For suppose  $\nabla_a$ is flat, and the pair $(\nabla_a, \phi)$ satisfies (R1), (R2), (R3). Let  $\psi$ be any smooth function (with the same domain as $\phi$) such that $\nabla^a \, \nabla^b \psi$ vanishes everywhere, but $\nabla^b\psi $ does not.\footnote{We can think of $\nabla^b\psi $ as the ``spatial gradient" of $\psi$. The stated conditions impose the requirement that $\nabla^b\psi $ be constant on all spacelike submanifolds (``time slices"), but  not vanish on all of them.}  If we set $\overline{\phi} = \phi + \psi$, and take $\overline{\nabla}_a$ to be the derivative operator relative  to which the action of $\stackrel{g \hspace{.4em}}{\nabla_a}$ is given by  $\overline{C}\hspace{.1em}^a_{\ bc}  = - t_{bc} \, \nabla^a \overline{\phi}$, then $\overline{\nabla}_a$ is flat and the pair $(\overline{\nabla}_a, \overline{\phi})$ satisfies conditions (R1),  (R2), (R3) as well.\nopagebreak[3]\footnote{It follows directly from the way $\overline{\nabla}_a$ was defined that the pair $(\overline{\nabla}_a, \overline{\phi})$ satisfies conditions (R1) and (R2). (The argument is almost exactly the same as that used in an earlier note to prove (G1) and (G2) in the Geometrization Theorem.) What must be shown that is that  $\overline{\nabla}_a$ is flat, and that the pair $(\overline{\nabla}_a, \overline{\phi})$ satisfies Poisson's equation:  $\overline{\nabla}{}^a \, \overline{\nabla}_a \, \overline{\phi} =  4  \, \pi  \, \rho$.
We do so by showing that  (i) $\overline{R}\hspace{.1em}^a_{\ bcd} \, = \,  R^a_{\ bcd}$, (ii) $\nabla^a \, \nabla_a  \, \psi =  \mathbf{0}$,  and  (iii) $ \overline{\nabla}{}^a \, \overline{\nabla}_a \, \alpha   =  \nabla^a \, \nabla_a \, \alpha$, for \emph{all} smooth scalar fields $\alpha$ on  $M$. (It follow immediately from (ii) and (iii) that 
$\overline{\nabla}{}^a \, \overline{\nabla}_a \, \overline{\phi} = \overline{\nabla}{}^a \, \overline{\nabla}_a \, \phi + \overline{\nabla}{}^a \, \overline{\nabla}_a \, \psi =  \nabla^a \, \nabla_a  \, \phi  +  \nabla^a \, \nabla_a  \, \psi =  4  \, \pi  \, \rho$.) 

We know from the uniqueness clause of (G2) in the Geometrization Theorem that the action of $\stackrel{g \hspace{.4em}}{\nabla_a}$ with respect to $\nabla_a$ is given by the field  $C^a_{\ bc}  =  -t_{bc} \, \nabla^a \phi$.  It follows that the action of $\overline{\nabla}_a$ relative to that of $\nabla_a$ is given by  $\widehat{C}\hspace{.1em}^a_{\ bc} = -\overline{C}\hspace{.1em}^a_{\ bc} + C^a_{\ bc} =  -t_{bc} \, \nabla^a  (-\overline{\phi} + \phi) =   t_{bc} \, \nabla^a  \psi$. So, arguing almost exactly as we did  in the proof of (G3) in the Geometrization Theorem, we have
\begin{equation} \label{non-uniqueness note 1}
\overline{R}\hspace{.1em}^a_{\ bcd} \, = \,  R^a_{\ bcd}  \, + \,  2 \, t_{b[d}  \nabla_{c]} \,  \nabla^a \psi. 
\end{equation}
Now it follows from $\nabla^a \, \nabla^b \psi = \mathbf{0}$ that
\begin{equation} \label{non-uniqueness note 2}
\nabla_c \, \nabla^a \psi  = t_c \, (\xi^n \nabla_n  \nabla^a \psi),
\end{equation}
where $t_{ab} = t_a t_b$, and  $\xi^n$ is any smooth future-directed unit timelike vector field on $M$.  Hence, $t_{b[d}  \nabla_{c]} \,  \nabla^a \psi = t_b \,  t_{[d} \,t_{c]} (\xi^n \nabla_n  \nabla^a \psi) = \mathbf{0}$.  This, together with (\ref{non-uniqueness note 1}), gives us (i).   And (ii) follows directly from (\ref{non-uniqueness note 2}). Finally, for (iii), notice that
\begin{eqnarray*}
\overline{\nabla}{}^a \, \overline{\nabla}_a \, \alpha   &=&  h^{ar}  \, \overline{\nabla}_r \, \overline{\nabla}_a \, \alpha  \, = \, h^{ar} \,  \overline{\nabla}_r \, \nabla_a  \, \alpha  = h^{ar} \, (\nabla_r \, \nabla_a \, \alpha + \widehat{C}\hspace{.1em}^n_{\ ra} \, \nabla_n \, \alpha ) \\ 
 {} &=& \nabla^a \, \nabla_a \, \alpha \, + \,  h^{ar} \,  t_{ra} \, (\nabla^n \, \psi) (\nabla_n \, \alpha) \, = \, \nabla^a \, \nabla_a \, \alpha. 
\end{eqnarray*}
 The final equality follows from the orthogonality condition.}
 \nopagebreak[3]
  \[
\begin{array}{rcl}
(\nabla_a, \phi)  &{} & {}  \\
\vspace{-.5em}
{} & \searrow &{}  \\
\vspace{-.5em}
{} & {} &\stackrel{g \hspace{.4em}}{\nabla_a} \\
\vspace{-.5em}
{} & \nearrow &{}  \\ 
\vspace{-.5em}
(\overline{\nabla}_a, \overline{\phi})  &{} & {}  
\end{array}
\] 

\vspace{.5em}

But, because $\nabla^b \, \psi$ is  non-vanishing (somewhere or other), the pairs $(\nabla_a, \phi)$ and  $(\overline{\nabla}_a, \overline{\phi})$ are distinct decompositions of $\stackrel{g \hspace{.4em}}{\nabla_a}$.   Relative to the first,  a point particle (with mass $m$  and four-velocity $\xi^a$) has acceleration $\xi^n \, \nabla_n \, \xi^a$ and is subject to a gravitational force $-m \, \nabla^a  \phi$.  Relative to the second, it has  acceleration $\xi^n \, \overline{\nabla}_r \, \xi^a = \xi^n \, \nabla_n  \, \xi^a - \nabla^a \, \psi$ and is subject to a gravitational force  $-m \, \overline{\nabla}{}^a \, \overline{\phi} = -m \, \nabla^a  \phi - m \, \nabla^a  \psi$.

As suggested at the beginning of the section, we can take this non uniqueness of recovery result to capture in precise mathematical language the standard claim that Newtonian gravitational force is a gauge quantity.  By the argument just given, if we can take the force on a point particle with mass $m$ to be  $-m \, \nabla^a  \phi$, we can equally well take it to be $-m \, \nabla^a  (\phi  +  \psi)$, where $\psi$ is any field satisfying $\nabla^a \, \nabla^b \psi = \mathbf{0}$.

\subsection{Recovering Global Geometric  Structure from ``Causal Structure"} \label{Recovering Geometric Structure}

There are many interesting and important issues concerning the  global structure of relativistic spacetimes that might be considered here --  the nature and significance of singularities,  the cosmic censorship hypothesis, the possibility of ``time travel", and others.\footnote{Earman \shortcite{Bangs} offers a comprehensive review of many of them.  (On the topic of singularities, I can also recommend  Curiel \shortcite{Curiel}.)} But we limit ourselves to a few remarks about one rather special  topic.  

In our discussion of relativistic spacetime structure, we started with geometric models $(M, g_{ab})$ exhibiting several levels of geometric structure,  and used the latter to define the (two-place) relations $\ll $  and  $ <$  on $M$.\footnote{Recall that $p \ll q$ holds if there is a future-directed timelike curve that runs from $p$ to $q$; and $p < q$ holds if there is a future-directed causal curve that runs from $p$ to $q$.}  The latter are naturally construed as  relations of ``causal connectibility (or accessibility)".  The question arises whether it is possible to work  backwards, i.e.,  \emph{start} with the pair $(M, \ll)$ or $(M, <)$,  with $M$ now construed as a bare point set, and recover the geometric structure with which we began. The question is  suggested by long standing interest on the part of some philosophers  in  ``causal theories" of time or spacetime.   It also figures centrally in a certain approach to quantum gravity developed by Rafael  Sorkin and co-workers. (See, e.g., Sorkin \shortcite{Sorkin-specimen,Sorkin-causalsets}.)

Here is one way to make the question precise. (For convenience, we work with the relation $ \ll $.)  

Let $(M, g_{ab})$ and $(\overline{M}, \overline{g}_{ab})$ be (temporally oriented) relativistic spacetimes. We say that a  bijection $\phi: M \rightarrow \overline{M}$ between their underlying point sets is a \emph{causal isomorphism} if, for all   $p$ and $q$ in $M$,
\begin{equation}
p \ll q   \ \  \Longleftrightarrow  \  \   \phi(p) \ll \phi(q).
\end{equation} 
Now we ask:  Does a causal isomorphism have to be a homeomorphism? a diffeomorphism? a conformal  isometry?\footnote{We know in advance that a causal isomorphism need not be a (full)  isometry because conformally equivalent metrics $g_{ab}$ and $ \Omega^2 \, g_{ab}$ on a manifold $M$ determine the same relation $\ll$.  The best one can ask for is that it be a conformal isometry, i.e., that it be a diffeomorphism that preserves the metric up to a conformal factor.}

Without further restrictions on  $(M, g_{ab})$ and   $(\overline{M}, \overline{g}_{ab})$, the answer is certainly `no' to all three questions.  Unless the ``causal structure"  (i.e., the structure determined by $\ll $) of a spacetime is reasonably well behaved, it provides no useful information at all. For example, let us say that a spacetime is \emph{causally degenerate} if $p \ll q$ for all points $p$ and $q$. \emph{Any} bijection between two causally degenerate spacetimes qualifies as a causal isomorphism.  But we can certainly find causally degenerate spacetimes whose underlying manifolds have different topologies (e.g., G\"odel spacetime and a rolled-up version of Minkowski spacetime).

There is a hierarchy of ``causality conditions" that is relevant here.  (See, e.g., Hawking and Ellis \shortcite[section 6.4]{Hawking-Ellis}.)  They impose, with varying degrees of stringency, the requirement that there exist no closed, or ``almost closed",  timelike curves. Here are three. 
\begin{enumerate}
\item [{}] \hspace{-2em} \emph{chronology}:  There do not exist closed timelike curves.  (Equivalently, for all $p$,  it is  \emph{not} the case that $p \ll p$.)
\item [{}]\hspace{-2em} \emph{future (resp.  past) distinguishability}:   For all points $p$, and all sufficiently small open sets $O$ containing $p$, no future directed (resp. past directed) timelike curve that starts at p, and leaves $O$, ever returns to $O$.  
\item [{}]\hspace{-2em}  \emph{strong causality}:    For all points $p$, and all sufficiently small open sets $O$ containing $p$, no future directed timelike curve that starts in $O$, and  leaves $O$,   ever returns to $O$.   
\end{enumerate}
It is clear that strong causality implies both future distinguishability and past distinguishability, and that each of the distinguishability conditions (alone) implies chronology.  Standard  examples (Hawking and Ellis \shortcite{Hawking-Ellis}) establish that the converse implications do not hold, and that neither distinguishability condition implies the  other. 

The names ``future distinguishability"  and ``past distinguishability" are  easily explained. For any $p$, let   $I^+(p)$ be the set $\{q: p \ll q\}$ and let  $I^-(p)$ be the set $\{q: q \ll p\}$.  Then  future distinguishability is equivalent to the requirement that,  for all $p$ and $q$, 
\begin{equation*}   
I^+(p) = I^+(q) \, \Rightarrow   \,  p=q.  
\end{equation*}   
And the counterpart requirement  with $I^+$ replaced by $I^-$ is equivalent to past distinguishability. 
  
We mention all this because it turns out that one gets a positive answer to all three questions above if one restricts attention to spacetimes that are \emph{both} future and past distinguishing.

\begin{proposition} \label{J Math Phys prop} 
Let $(M, g_{ab})$ and   $(\overline{M}, \overline{g}_{ab})$ be (temporally oriented) relativistic spacetimes that are past and future distinguishing, and let $\phi: M \rightarrow \overline{M}$ be a causal isomorphism.  Then  $\phi$ is a diffeomorphism and preserves $ g_{ab}$ up to a conformal factor, i.e., $\phi_{\star}g_{ab}$ is conformally equivalent to $\overline{g}_{ab}$.    
\end{proposition}

A proof is given in Malament \shortcite {Malament/causalrecovery}. A counterexample given there also shows that the proposition fails if the  initial restriction on causal structure is weakened to past distinguishability or to future distinguishability alone.

\newpage
\bibliographystyle{named}
\bibliography{GRSurveybibliography}


\end{document}